\documentclass[12pt]{article}

\usepackage{graphicx,bm,amssymb,amsmath}

\newcommand{\bi}{\begin{itemize}}
\newcommand{\ei}{\end{itemize}}
\newcommand{\ben}{\begin{enumerate}}
\newcommand{\een}{\end{enumerate}}
\newcommand{\be}{\begin{equation}}
\newcommand{\ee}{\end{equation}}
\newcommand{\bea}{\begin{eqnarray}}
\newcommand{\eea}{\end{eqnarray}}
\newcommand{\bc}{\begin{center}}
\newcommand{\ec}{\end{center}}
\newcommand{\ie}{{\it i.e.\ }}

\newcommand{\etal}{{\it et al.\ }}

\newcommand{\smallfrac}[2]{{\textstyle\frac{#1}{#2}}}
\newcommand{\half}{\smallfrac{1}{2}}

\newcommand{\tbox}[1]{{\mbox{\tiny #1}}}
\newcommand{\mbf}[1]{{\mathbf #1}}
\newcommand{\ofr}{(\mbf{r})}
\newcommand{\ofkr}{(k,\mbf{r})}
\newcommand{\ofa}{(\mbf{r}_1)}
\newcommand{\ofb}{(\mbf{r}_2)}
\newcommand{\drr}{d\mbf{r}_1 d\mbf{r}_2}
\newcommand{\dab}{|\mbf{r}_1 - \mbf{r}_2|}

\DeclareMathOperator{\vol}{vol}
\newcommand{\Ael}[2]{\ensuremath{\langle \phi_{#1}, \hat{A} \phi_{#2}\rangle}}
\newcommand{\Abar}{\overline{A}}
\newcommand{\ccla}{\tilde{C}_A}

\newcommand{\cclaw}{\ccla(\omega)}

\newcommand{\osm}{\omega_\tbox{sm}}
\newcommand{\arw}{a_\tbox{RW}}
\newcommand{\afp}{a_\tbox{FP}}
\newcommand{\abf}{a_\tbox{BF}}
\newcommand{\abfh}{a_\tbox{BF}^{(1/2)}}
\newcommand{\gbf}{\gamma_\tbox{BF}}

\newcommand{\gdes}{\Gamma_\tbox{desym}}

\newcommand{\nsc}{{N_\tbox{sc}}}
\newcommand{\ep}{\epsilon}

\newcommand{\ds}{{d\mbf{s}}}    
\newcommand{\dr}{{d\mbf{r}}}    
\newcommand{\cp}{Conjecture~\ref{conj:power}}             
\newcommand{\cd}{Conjecture~\ref{conj:diag}}
\newcommand{\cq}{Conjecture~\ref{conj:que}}
\newcommand{\co}{Conjecture~\ref{conj:offd}}
\newtheorem{thm}{Theorem}
\newtheorem{conj}{Conjecture}
\newtheorem{rmk}{Remark}[section]
\newtheorem{lem}{Lemma}[section]

\title{Asymptotic rate of quantum ergodicity in chaotic Euclidean billiards}
\author{Alex H. Barnett\\
{\normalsize \em Courant Institute, New York University,
251 Mercer St, New York, NY 10012}\footnote{now at
Department of Mathematics, Dartmouth College,
6188 Bradley Hall, Hanover, NH, 03755. {\small \tt ahb@math.dartmouth.edu}}}

\date{\today}

\begin{document}
\maketitle

\begin{abstract}
The Quantum Unique Ergodicity (QUE) conjecture of Rudnick-Sarnak is
that every
eigenfunction $\phi_n$ of the Laplacian on a manifold with
uniformly-hyperbolic geodesic flow
becomes equidistributed
in the semiclassical limit (eigenvalue $E_n \to \infty$),
that is, `strong scars' are absent.
%
%
We study numerically
the {\em rate} of equidistribution
for a uniformly-hyperbolic Sinai-type planar Euclidean billiard
with Dirichlet boundary condition (the `drum problem')
at unprecedented high $E$ and statistical accuracy,
via the matrix elements $\Ael{n}{m}$
of a piecewise-constant test function $A$.
By collecting $30000$ diagonal elements
(up to level $n\approx 7\times 10^5$)
we find that their variance decays with eigenvalue as a
power $0.48 \pm 0.01$, close to the 
estimate $1/2$ of Feingold-Peres (FP).
This contrasts the results of existing studies, which have been limited to 
$E_n$ a factor $10^2$ smaller.
We find strong evidence for QUE in this system.
%
%
We also compare off-diagonal variance, as a function of
distance from the diagonal, against FP 
at the highest accuracy ($0.7\%$) thus far in any chaotic system.
We outline the efficient scaling method 
used to calculate eigenfunctions.
\end{abstract}

\section{Introduction}

The nature of the quantum (wave) mechanics of Hamiltonian
systems whose classical
counterparts are chaotic has been of long-standing
interest, dating back to Einstein in 1917
(see~\cite{history} for a historical
account).
The field now called `quantum chaos' is
the study of such quantized systems in the
short wavelength (semiclassical, $\hbar\rightarrow0$ or high energy) limit,
and has become
a fruitful area of enquiry for both physicists
(for reviews see \cite{gutz,hellerhouches}) and mathematicians
(see \cite{sarnakCRM,baltimore,zencyc})
in recent decades.
In contrast to those of
integrable classical systems, eigenfunctions are irregular.
A central issue, and the topic of this numerical study,
is their behavior in the semiclassical limit.

We consider billiards, a paradigm problem in this field.
A point particle is
trapped inside a bounded planar domain $\Omega \subset \mathbb{R}^2$
and bounces elastically off the boundary $\Gamma=\partial\Omega$.
Its phase space coordinate is
$(\mbf{r},\theta) \in \Omega \times S^1$,
with position $\mbf{r}:=(x,y)$ and (momentum) direction  $\theta$.
The corresponding quantum-mechanical system is the spectral
problem of the Laplacian in $\Omega$ with homogeneous local
boundary conditions which we may (and will) take to be Dirichlet,
\bea
	-\Delta \phi_n &=& E_n \phi_n,
\label{eq:helm}
	\\
	\phi_n(\Gamma)&=&0.
\label{eq:dir}
\eea
Eigenfunctions $\phi_n$ are real-valued and normalized
\be
	\langle \phi_n, \phi_m\rangle \; := \;
	\int_\Omega \phi_n\ofr \phi_m\ofr \dr
	\;=\;\delta_{nm},
\label{eq:on}
\ee
where $\dr:=dxdy$ is the usual area element,
and the corresponding `energy' (or frequency)
eigenvalues are ordered
$E_1 < E_2 \le E_3 \le \cdots \infty$.
We will also write $E_j = k_j^2$ where $k_j$ is the wavenumber.
This Dirichlet eigenproblem~\cite{KS}, also known as the
membrane or drum problem, has a rich 150-year history of applications to
acoustics, electromagnetism, optics, vibrations, and quantum mechanics.

\begin{figure}
\includegraphics[width=\textwidth]{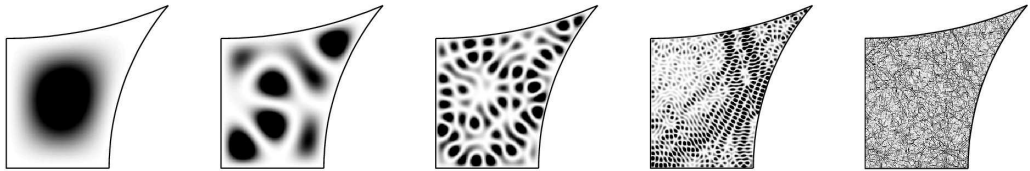}
\caption{\label{fig:q}
\small
Illustration of spatial equidistribution of $|\phi_n|^2$ (shown as
density plots; larger values are darker)
with increasing eigenvalue:
$n^{th}$ Dirichlet eigenfunction for mode numbers
$n=1,10,100,1000$ and $n\approx 50000$.}
\end{figure}

When the classical dynamics (flow) is ergodic it is well-known that
almost all eigenfunctions are `quantum ergodic', in the following sense.
%
\begin{thm}[Quantum Ergodicity Theorem (QET)~\cite{schnir,zel,cdv,zzw}]
Let $\Omega \in \mathbb{R}^2$ be a 2D compact domain with piecewise
smooth boundary whose classical flow is ergodic.
Then for all $n$ except a subsequence of vanishing density,
\be
	\Ael{n}{n} - \Abar \to 0
	\qquad \mbox{as} \;n\to\infty,
\ee
for all well-behaved functions $A:\Omega \to\mathbb{R}$.
\label{thm:qet}
\end{thm}
%
The operator $\hat{A}$ is multiplication by $A$,
a `test function' whose spatial average is
$\Abar := \frac{1}{\vol(\Omega)} \int_\Omega A\ofr \,\dr$.
The physical interpretation is that
almost all quantum expectation
values (diagonal matrix elements) of the observable $\hat{A}$
tend to their classical expectation $\Abar$,
an example of the Correspondence Principle of quantum mechanics.
Note the choice `almost all' need not depend on $A$~\cite{zel}.
Equivalently, almost all probability densities $|\phi_j|^2$ tend
to the uniform function $\frac{1}{\vol(\Omega)}$, weakly in $L^1(\Omega)$
(see Fig.~\ref{fig:q}).
The proof relies on the machinery of semiclassical analysis including
Fourier Integral Operators (see
\cite{zencyc,martinez} and references within).
In our context the `well-behaved' requirement is, loosely speaking,
that $A$ not be oscillatory on the (vanishing) wavelength scale $1/k_j$.
(Formally $\hat{A}$ must be a zeroth-order pseudo-differential operator,
a Weyl quantization of the principal symbol $A\ofr$).
We investigate only multiplication operators, that is, no
dependence on momentum, and, as we will see below,
by further restricting to piecewise constant test functions
we will exploit a huge numerical efficiency gain.

Our study is motivated by the fact that QET tells us nothing about
the rate of convergence of $\Ael{n}{n}$ or
the density of the excluded subsequence,
both of which are needed to understand the practical applicability
of the QET in quantum or other eigenmode systems.
We are interested in how the size of the deviation $\Ael{n}{n} - \Abar$
varies with eigenvalue $E_n$.
We define its `local variance' (mean square value) at energy $E$ by
\be
	V_A(E)
	:=
	\frac{1}{N_L(E)}\sum_{n:E_n \in [E,E+L(E)]} 
	\left|\Ael{n}{n} - \Abar\right|^2,
\label{eq:vae}
\ee
where $N_L(E):=N(E+L(E))-N(E)$ and the level counting function is
\be
N(E):=\#\{n:E_n \le E\}
\label{eq:ne}
\ee
Here we envisage an energy window width $L(E) = O(E^{1/2})$,
that is, a wavenumber window of width $O(1)$: this
contains $O(E^{1/2})$ eigenvalues by Weyl's Law~\cite{gutz,zencyc}.
For practical reasons (Section~\ref{sec:diag})
we will in fact use other $L(E)$ widths which nevertheless
contain many ($\approx10^3$) eigenvalues.
We will test the following asymptotic form for the variance.
\begin{conj}[Power-law diagonal variance]
For ergodic flow, as $E\to\infty$, there is the asymptotic form,
for some $a$ and $\gamma$,
\be
V_A(E) \sim a E^{-\gamma}.
\ee
\label{conj:power}
\end{conj}
A random-wave model for eigenfunctions
(Section~\ref{sec:rw}) predicts $\gamma=1/2$ and a certain prefactor $\arw$.
We will also test a more elaborate heuristic
from the physics literature (Section~\ref{sec:fp})
which predicts $\gamma=1/2$ and the following different prefactor.
\begin{conj}[Feingold-Peres diagonal variance~\cite{fp86,EFKAMM}]
For ergodic flow with no symmetries other than time-reversal,
\be
V_A(E) \sim \frac{g\tilde{C}_A(0)}{\vol(\Omega)}  E^{-1/2}
\ee
where the symmetry factor is $g=2$.
\label{conj:diag}
\end{conj}
Here the prefactor $\tilde{C}_A(\omega)$ is the
power spectral density (also known as classical variance~\cite{baltimore},
spectral measure~\cite{zqmix,zencyc},
or fluctuations intensity~\cite{doronfrc}),
\be
	\cclaw \; :=\;
	\int_{-\infty}^\infty C_A(\tau) e^{i\omega\tau} d\tau,
\label{eq:cw}
\ee
the time autocorrelation of $A$ being 
\be
	C_A(\tau) :=
	\lim_{T\to\infty} \frac{1}{T}\int_{0}^{T}
	\!A(\mbf{r}(t)) A(\mbf{r}(t+\tau)) \, dt ,
\label{eq:ctraj}
\ee
where $\mbf{r}(t)$ is any uniformly-distributed (ergodic) trajectory.
%
\begin{rmk}
In the physics literature much stronger conjectures are often discussed,
such as {\em individual}
matrix elements $\Ael{n}{n}$ being pseudo-randomly
distributed with variance given as above. We present evidence for this
in Section \ref{sec:que}.
At the other extreme, proven theorems
(and some numerical work~\cite{baecker}) often involve
sums of the form $S_p(E;A) :=
N(E)^{-1} \sum_{j:E_j\le E} \left|\Ael{j}{j} - \Abar\right|^p$.
Note that asymptotic decay $S_p(E;A) \sim bE^{-p\gamma/2}$ is equivalent to
\cp.
However we will study $V_A(E)$ rather than $S_p(E;A)$ for the following
important practical reasons:
\ben
\item Narrow spectral windows are needed rather than the complete
spectrum, allowing
much higher eigenvalues to be included in the statistics.
\item Asymptotic behavior will emerge sooner since data at high $E_n$
are not averaged with that from the lower part of the spectrum.
%
%
\een
\label{rmk}
\end{rmk}
About 10\% agreement\footnote{These
researchers also studied the hydrogen atom in a strong magnetic
field (nearly completely ergodic with sticky islands),
and found some agreement
at the 20\% level, however they admit that the agreement
was `unexpectedly good' since it depended on a choice of smoothing
parameter.}
with \cd\  has been shown
in the quantum bakers map~\cite{EFKAMM}, and rough agreement
with $\gamma=1/2$ has been found in hyperbolic polygons~\cite{aurich}.
For general ergodic flow,
proven bounds on the power-law rate $\gamma$ in \cp\  are quite wide:
Zelditch~\cite{zelupper} has shown that $S_p(E;A) = O((\log E)^{-p/2})$,
implying $\gamma > 0$,
and shown~\cite{zellower}
that for generic $A$ (more precisely, one with nonzero mean
sub-principal symbol), $\gamma \le 1$.

\begin{figure}[tpb]
\bc
\includegraphics[width=\textwidth]{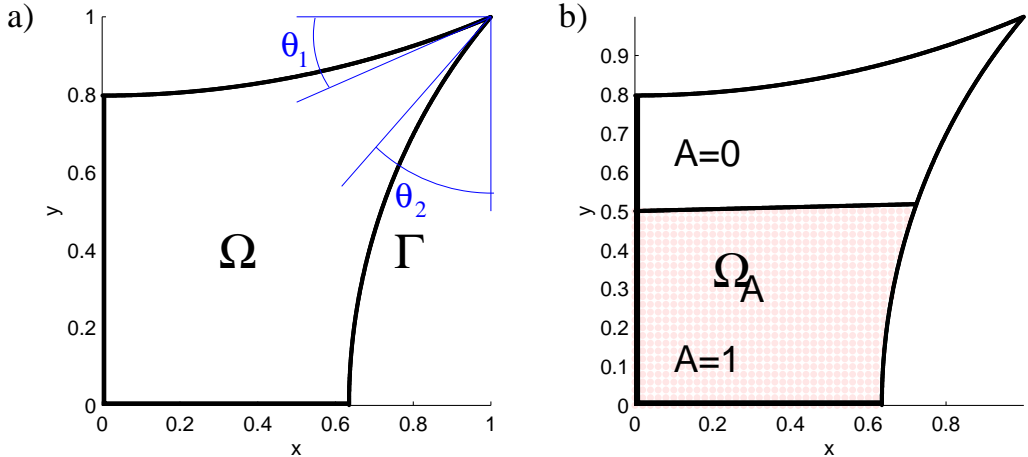}
\ec
\caption{Billiard and test function used in this study.
a) Quarter generalized Sinai billiard formed by two circular arcs
which meet at $(1,1)$ at angles $\theta_1 = 0.4$ to the horizontal
and $\theta_2 = 0.7$ to the vertical. The straight sections lie on
the axes and meet the arcs at right angles.
b) Piecewise-constant $A\ofr$ which takes the value $1$ inside the
region $\Omega_A$ and zero elsewhere. $\partial\Omega_A\setminus\Gamma$ is
a straight line which does not meet either wall at right angles.
\label{fig:bil}
}
\end{figure}

There are special `arithmetic' manifolds with ergodic flow
for which powerful number-theoretic tools~\cite{sarnak95,baltimore}
allow much more to be proven.
For the quotient manifold $\mbf{H}^2/SL(2,\mathbb{Z})$
Luo and Sarnak~\cite{lsquad} recently showed
$S_2(E;A) \sim B(A)E^{-1/2}$, where the prefactor is a quadratic
form $B(\cdot)$ diagonalized by the eigenfunctions
themselves. It takes the value $B(\phi_n) = \tilde{C}(0)L(\half,\phi_n)$,
where $L(\cdot,\cdot)$ is an L-function~\cite{baltimore}.
Thus the power law appearing in \cd\  holds
but the prefactor differs from $g=2$.
This is hardly surprising;
Sarnak~\cite{baltimore} notes that a simple reflection symmetry is enough
to cause $g=0$.
Arithmetic systems are very special and have many symmetries:
all periodic orbits
possess a single Lyapunov exponent, and eigenvalue spacing
statistics are unusual for ergodic systems~\cite{bogarith}.
This makes the study of a planar Euclidean billiard
system {\em without} symmetry (where Lyapunov exponents differ),
for which no number-theoretic analytic tools exist,
particularly interesting.
We choose such a generic billiard (with hyperbolic, \ie Anosov,
flow~\cite{disperse})
for our numerical experiments (see Fig.~\ref{fig:bil}).
Numerical tests of
\cp\  (in the form of $S_1(E;A)$, see Remark~\ref{rmk}) exist
for low eigenvalues ($n<6000$) in the Anosov cardioid billiard~\cite{baecker}:
various powers were found in the range $\gamma=0.37$ to 0.5,
and up to 20\% deviations from the prefactor in \cd.

We address two other questions, the first regarding the
excluded subsequence in Theorem~\ref{thm:qet}, the second regarding
off-diagonal matrix elements.
\begin{conj}[Quantum Unique Ergodicity (QUE)~\cite{RS}]
There is no excluded subsequence in Thm.~\ref{thm:qet}.
\label{conj:que}
\end{conj}
`Unique' refers to the existence of only one `quantum limit'
(any measure to which $|\phi_n|^2$ tends weakly).
Made in the context of general negatively-curved manifolds,
this conjecture has
remarkably been proved for arithmetic manifolds~\cite{linden}.
In contrast, QUE has been proven {\em not} to hold for certain-dimensional
quantizations of Arnold's cat map~\cite{faure}.
This begs the question: for which ergodic billiards, if any,
does this conjecture hold?
For instance there is strong evidence~\cite{tanner,baeckerbb}
(but no proof~\cite{donnelly})
that a sequence of `bouncing ball'
modes (eigenfunctions concentrated on cylindrical or neutrally stable
orbits which occupy zero measure in phase space)
persists to arbitrarily high eigenvalues
in ergodic billiards such as Bunimovich's stadium.
Since such modes are not spatially
uniform, the conjecture would not hold for this shape.
There are also more subtle nonuniformity effects:
in the physics community enhancements of $|\phi_n|^2$, dubbed `scars' by
Heller~\cite{hel84,hellerhouches},
are known to exist along isolated unstable periodic orbits,
and there has been a long-standing
debate as to whether these may prevent QUE.
In Section~\ref{sec:scar} we discuss how our results relate to scarring.

Finally we consider the size of off-diagonal matrix elements $\Ael{n}{m}$.
Define $\Delta_k(E):= 2\pi/[E^{1/2}\vol(\Omega)]$ as the mean level
spacing (in wavenumber).
\begin{conj}[Feingold-Peres off-diagonal variance~\cite{fp86,wilk87}]
Fix $\omega\in\mathbb{R}$. Then for ergodic flow, as $E\to\infty$ there is
the asymptotic result,
\be
V_A(E;\omega) \;:=\; \frac{\Delta_k(E)}{2\epsilon(E)\,N_L(E)}
\sum_{   \substack{m,n:E_n \in [E,E+L(E)]\\
         |k_m-k_n-\omega|\le\epsilon(E)}}
	\!\!\!|\Ael{n}{m}|^2
\;\sim\; \frac{\tilde{C}_A(\omega)}{\vol(\Omega)} E^{-1/2}
\label{eq:fp}
\ee
where $0 < \epsilon(E) = O(E^{-1/4})$.
\label{conj:offd}
\end{conj}
$V_A(E;\omega)$ measures the mean
off-diagonal quantum variance a `distance' (in wavenumber units) $\omega$
from the diagonal, which is thus given by
classical variance at frequency $\omega$
(see Section~\ref{sec:fp} for a heuristic argument).
As above, the choice $L(E)=O(E^{1/2})$ is envisaged. The
rate at which the wavenumber window $\epsilon(E)$ vanishes
includes a growing number $O(E^{1/4})$ of modes.
The result (with equivalent choices of $L(E)$ and $\epsilon(E)$) has
been proved for
Schr\"{o}dinger operators with smooth confining potential using coherent
states~\cite{combe2}.
\co\ is a stronger version of the spectral measure theorem~\cite{zqmix,tate}
\be
\lim_{E\to\infty}\frac{1}{N(E)} \sum_{
  \substack{m,n:E_n \in [0,E]\\
    \alpha< k_m-k_n <\beta}
}
\!\!\!|\Ael{n}{m}|^2
\; = \;
\int_\alpha^\beta \frac{\tilde{C}_A(\omega)}{2\pi} d\omega,
\ee
which holds for $\alpha<\beta$ independent of ergodicity.
It is known $V_A(E;\omega)$ vanishes for ergodic weak-mixing flows
(weak-mixing ensures $\tilde{C}_A(\omega)$ is a bounded function),
but without any proven rate~\cite{zeloffdiag}.
The most accurate previous numerical test of
\co\
is the 10\% agreement found for billiards
with $\hat{A}$ a singular boundary operator~\cite{dil}.
Other tests have included the quartic oscillator~\cite{austin}, bakers
map~\cite{boose}, and lima\c{c}on
billiard~\cite{prosen}.


Existing numerical studies of all the above conjectures
share the features of low and poorly-quantified accuracy
(\ie lack of statistical rigor in the tests),
and relatively low mode numbers ($n\sim 10^3$ to $10^4$).
In this work we remedy both these flaws by performing a large-scale
study using non-standard
cutting-edge numerical techniques which excel at very high eigenvalues.
In Section \ref{sec:fp} we review heuristic arguments for
Conjectures~\ref{conj:power}, \ref{conj:diag} and \ref{conj:offd}.
Then in Section \ref{sec:num} (which refer to the three Appendices)
we outline numerical methods;
we emphasize there are several innovations.
However the reader interested in the results on the four Conjectures,
and discussion, may
skip directly to Section \ref{sec:res}.
We summarize our conclusions in Section~\ref{sec:conc}.

\begin{figure}[!tpb]
\bc
\includegraphics[width=\textwidth]{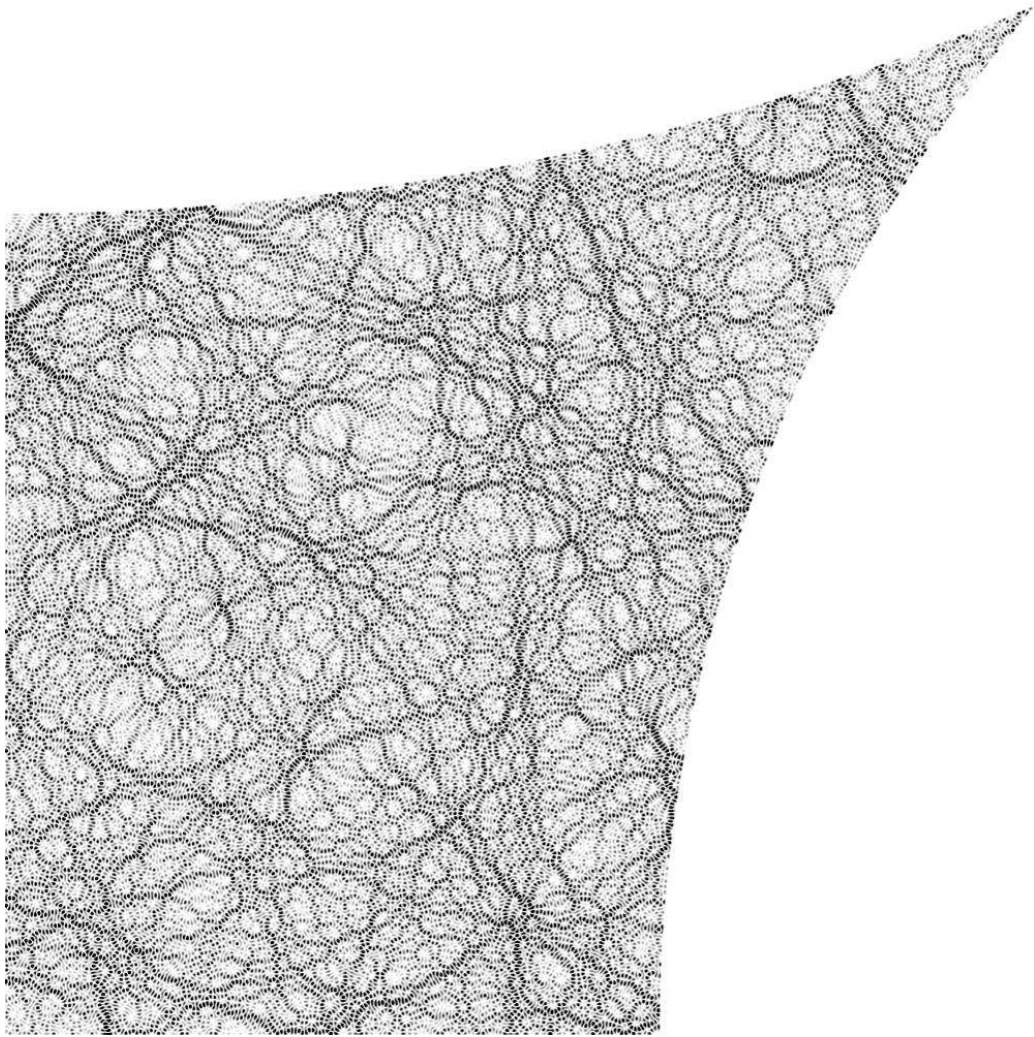} 
\ec
\caption{Density plot of $|\phi_n|^2$ for an eigenfunction
with $k_n = 999.90598\cdots$, that is,
$E_n \approx 10^6$, and level number $n \approx 5\times 10^4$.
There are about 225 wavelengths across the
diagonal. 
\label{fig:efunc}
}
\end{figure}
\begin{figure}[tpb]
\bc
\includegraphics[width=\textwidth]{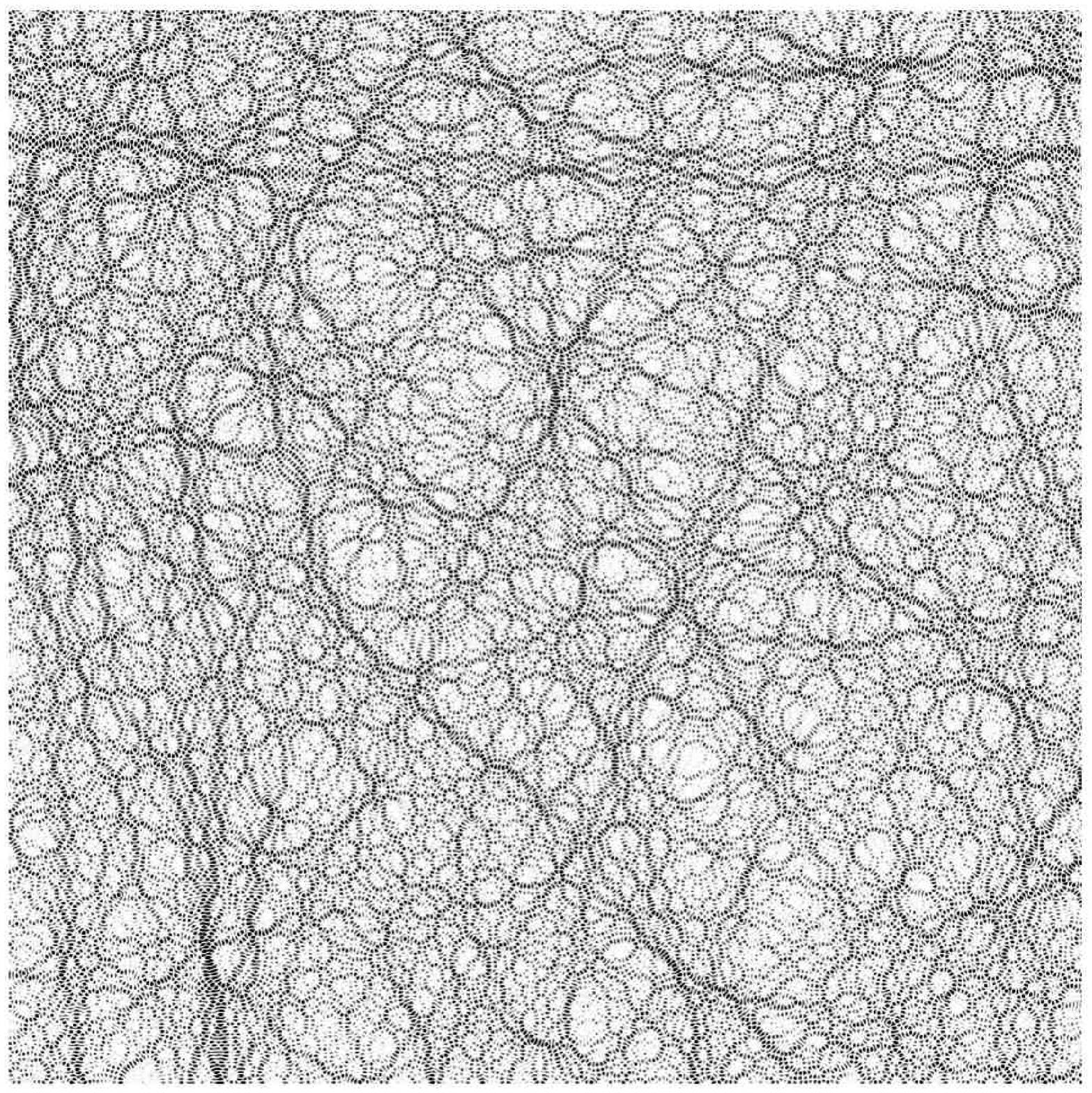}   
\ec
\caption{Density plot of one sample
from the ensemble of random plane waves with the same 
wavenumber magnitude $|\mbf{k}| = k$
and mean intensity as the eigenfunction in Fig.~\ref{fig:efunc},
shown in a square region of space.
(Note there are no boundary conditions imposed).
The `stringy' structures visible to
the eye are a feature of the constant wavevector magnitude;
they disappear if a finite range of $k$ is included~\cite{hellerhouches}.
\label{fig:ranpw}
}
\end{figure}

\section{Heuristic arguments for ergodicity rate}
\label{sec:sc}

Here we review some arguments for Conjectures~\ref{conj:power},
\ref{conj:diag} and \ref{conj:offd} from the physics literature.
We feel these are
appropriate since quantum chaos is somewhat of a cross-over area
between mathematics and physics.

\subsection{Random wave model}
\label{sec:rw}

Berry~\cite{berry77} put forward the conjecture that chaotic
eigenfunctions should look locally like a
superposition of plane waves of fixed energy $E$, traveling in all directions,
with random amplitudes and phases (compare Figs.~\ref{fig:efunc} and
\ref{fig:ranpw}).
We will show (following~\cite{EFKAMM}) that this model satisfies
\cp\  with certain $\gamma$ and $a$.
We note that Zelditch has more rigorously considered models
of random orthonormal bases on manifolds,
and shown that they possess similar ergodicity properties to
quantum chaotic eigenfunctions~\cite{zonb1,zonb2}.

Using the notation $\langle\,\cdot\,\rangle$ to denote averaging over
the ensemble of plane wave coefficients, the
two-point correlation of a random-wave field is~\cite{berry77}
\be
	\langle \phi(\mbf{r}_1) \phi(\mbf{r}_2) \rangle
	=
	\frac{1}{\vol(\Omega)} J_0(k\dab) ,
\label{eq:2pt}
\ee
where we use the normalization $\vol(\Omega) \langle \phi^2\rangle = 1$
appropriate to the billiard area. This applies to
both real (time-reversal symmetric) and complex (non time-reversal symmetric)
waves; from now we stick to the real case.
Because the model is a statistical one, it is meaningful to speak of the
variance of a particular matrix element within the ensemble,
namely $\mbox{var}(A_{nm})$ where we use $A_{nm}:=\Ael{n}{m}$.
A simple calculation using
(\ref{eq:2pt}) and Wick's theorem for Gaussian random variables gives
\be
	\langle \phi_n\ofa \phi_m\ofa \phi_n\ofb \phi_m\ofb \rangle
	=
	\frac{1}{\vol(\Omega)^2} \left[ \delta_{nm} + g_{nm} J_0^2(k\dab) \right]
\label{eq:22pt}
\ee
where
\be
	g_{nm}
	:=
	\left\{\begin{array}{ll}g,&n=m,\\1,&n\neq m.\end{array} \right.
\ee
Here the symmetry factor is $g=2$; we will shortly see it gives
the ratio of diagonal to off-diagonal variance.
A key assumption made was that $\phi_n$ and $\phi_m$
are statistically {\em independent}
members of the ensemble when $n\neq m$,
a natural one in the context of random matrix theory (RMT)
($g=2$ for time-reversal symmetry
is a standard result in the Gaussian Orthogonal Ensemble~\cite{RMT}).
The independence assumption, like the random wave model itself,
remains a heuristic one, albeit one with numerical support.
Clearly orthogonality (\ref{eq:on}) dictates that, when
considered as functions over all of $\Omega$, $\phi_n$ and $\phi_m$
cannot be independent!
(As was mentioned there are improved models which take this into
account~\cite{zonb2}).
An intuitive argument can be made if a restriction to a small subregion
of $\Omega$ is made: the eigenfunctions behave like random orthogonal
vectors, and projections of such
vectors onto a much smaller-dimensional subspace become
approximately independent.

Now we use (\ref{eq:22pt}) to evaluate the variance
diagonal and off-diagonal matrix elements, writing the variance
as the mean square minus the square mean,
\bea
	\mbox{var} (A_{nm})
	&=&
	\langle \left| \int_\Omega \phi_n\ofr \phi_m\ofr A\ofr \dr
	\right| ^2 \rangle  \; - \; \langle A_{nm} \rangle ^2 \nonumber \\
	&=&
	\frac{g_{nm}}{\vol(\Omega)^2} \int_\Omega\int_\Omega
	J_0(k_n\dab) J_0(k_m\dab) \drr \nonumber \\
	&\approx&
	\frac{g_{nm}}{\pi\vol(\Omega)} E^{-1/2} \int_\Omega\int_\Omega
	\frac{A\ofa A\ofb}{\dab}\drr,
\label{eq:varanm}
\eea
where we used $\langle A_{nm} \rangle = \delta_{nm} \Abar$. 
In the final step two approximations have been made: i)
$L|k_n-k_m| \ll 1$ where $L$ is the largest spatial scale of $A\ofr$,
meaning that the two Bessel functions always remain in phase so can be set
equal,
and
ii) the asymptotic form
$J_0(x) \sim (2/\pi x)^{1/2}\cos(x - \pi/4)$ was used, and $\cos^2$ replaced
by its average value $\half$, giving a semiclassical expression
valid when $kl \gg 1$, where $l$ is the smallest relevant spatial scale
in $A\ofr$.
Considering the diagonal and off-diagonal cases, (\ref{eq:varanm}) implies
\be
	\mbox{var} (A_{nn}) = g \, \mbox{var} (A_{nm})
\label{eq:symm}
\ee
in a region $n\approx m$ close enough to the diagonal.
The diagonal case of (\ref{eq:varanm}) gives the power law $\gamma=1/2$,
\be
	V_A(E) \; \approx \; \arw E^{-1/2},
\label{eq:rwg}
\ee
where the prefactor takes the form of a Coulomb interaction energy of
the `charge density' $A\ofr$,
\be
	\arw = \frac{g}{\pi\vol(\Omega)}  \iint \frac{A\ofa A\ofb}{\dab}\drr.
\label{eq:rwa}
\ee
Note that this model takes no account of the billiard shape
or boundary conditions.

\subsection{Classical autocorrelation argument (FP)}
\label{sec:fp}

Feingold and Peres~\cite{fp86} were the first to derive a
semiclassical
expression for diagonal variance in chaotic systems. There are two steps:
\ben
\item [(i)] relating off-diagonal variance to $\cclaw$ the power spectral
density, yielding \co, then
\item [(ii)] relating diagonal variance to off-diagonal variance
close to the diagonal, yielding \cd.
\een

{\bf Step (i)}:
Our presentation is loosely based on Cohen~\cite{doronfrc}; we emphasize that
this is not a mathematical proof, rather one form of a heuristic common in
physics literature (cf.\ \cite{wilk87,prosen,hort,EFKAMM}).
The autocorrelation (\ref{eq:ctraj})
of the `signal' $A(\mbf{r}(t))$ associated
with a uniformly-distributed ergodic unit-speed trajectory $\mbf{r}(t)$
is $C_A(\tau) = \overline{A(0) A(\tau)}$,
where the ergodic theorem was used to rewrite the time average
as an average over initial phase space locations $(\mbf{r}_0, \theta_0)$.
Fixing $\tau$ gives the function of phase space
${\cal A}(\mbf{r}_0,\theta_0) := A(0)A(\tau)$.
Applying QET to this function ${\cal A}$, gives as $E\to\infty$,
\be
        N_L(E)^{-1}\!\!\!\sum_{n:E_n\in[E,E+L(E)]}
	\langle\phi_n, \hat{A}(0) \hat{A}(\tau) \phi_n\rangle
	\;\sim\; C_A(\tau),
\label{eq:qm}
\ee
where $\hat{A}(t)$ is the quantization of $A$, shifted in time
(according to the Heisenberg picture of quantum mechanics).
A window $L(E) = O(E^{1/2})$ is sufficient for validity of
QET~\cite{zencyc}.
Assuming a wave dispersion relation $\omega = k$ the operator
$\hat{A}(t)$ is expressed in the eigenfunction basis,
\be
	\langle \phi_n, \hat{A}(t) \phi_m\rangle \; = \;
	A_{nm} e^{-i(k_m-k_n)t}.
\ee
Using this and inserting a sum over projections onto all eigenfunctions
into (\ref{eq:qm}) gives
\be
	N_L(E)^{-1}\!\!\!\sum_{n:E_n\in[E,E+L(E)]}\,
	\sum_{m=1}^\infty
	|A_{nm}|^2 e^{-i (k_m-k_n)\tau} \; \sim \; C_A(\tau).
\label{eq:ct}
\ee
Taking the inverse Fourier transform of the definition in
(\ref{eq:fp}),
recognising $\lim_{\epsilon\to0}(2\epsilon)^{-1}1_{[-\epsilon,\epsilon]}$
as the Dirac delta function, and using (\ref{eq:ct}), gives
\be
	\frac{1}{2\pi} \int e^{-i\omega\tau} V_A(E;\omega)\, d\omega
	\; \sim \; \frac{\Delta_k(E)}{2\pi} C_A(\tau).
\ee
Finally, taking a Fourier transform and substituting $\Delta_k(E)$
gives \co.

{\bf Step (ii)}:
To approach the diagonal we take the limit $\omega\to 0$;
here $\cclaw$ is well-defined and bounded since the flow is weak-mixing
\cite{zqmix}.
The expectation that diagonal variance should exceed off-diagonal variance
by the time-reversal invariance symmetry factor (\ref{eq:symm}) with $g=2$
gives \cd.
Feingold-Peres~\cite{fp86} justify this by considering
$\phi_{\pm}:=(\phi_n\pm\phi_m)/\sqrt{2}$ for eigenfunctions $\phi_n$ and
$\phi_m$ with sufficiently small $E_n-E_m$.
Matrix elements are then
$\Ael{+}{-} = (A_{nn} - A_{mm})/2$, where $A=A^{*}$ was used;
these are expected (by rotational invariance)
to have the same variance as $A_{mn}$.
Treating these quantities as {\em independent}
statistical variables, taking
variances, and recognising that mean values $\Abar$ cancel
on the right, gives
$\mbox{var}(A_{mn}) = \mbox{var}(A_{+-}) = \mbox{var}(A_{nn})/2$,
that is, $g=2$.
Although this was not discussed by FP, the system must be assumed to be
without further symmetry, or other values of $g$ may result.

Finally,
the physical significance of the correlation functions used above should be
noted: the right-hand side of (\ref{eq:fp}) is proportional to
dissipation (heating) rate in a classical system driven at frequency
$\omega$ with the forcing function $A$, and the left-hand side
to quantum dissipation rate under equivalent forcing (with $\hat{A}$)
within linear response theory (for reviews see~\cite{doronfrc,wlf,mythesis}).

\begin{figure}[tbp]
\bc
\includegraphics[width=\textwidth]{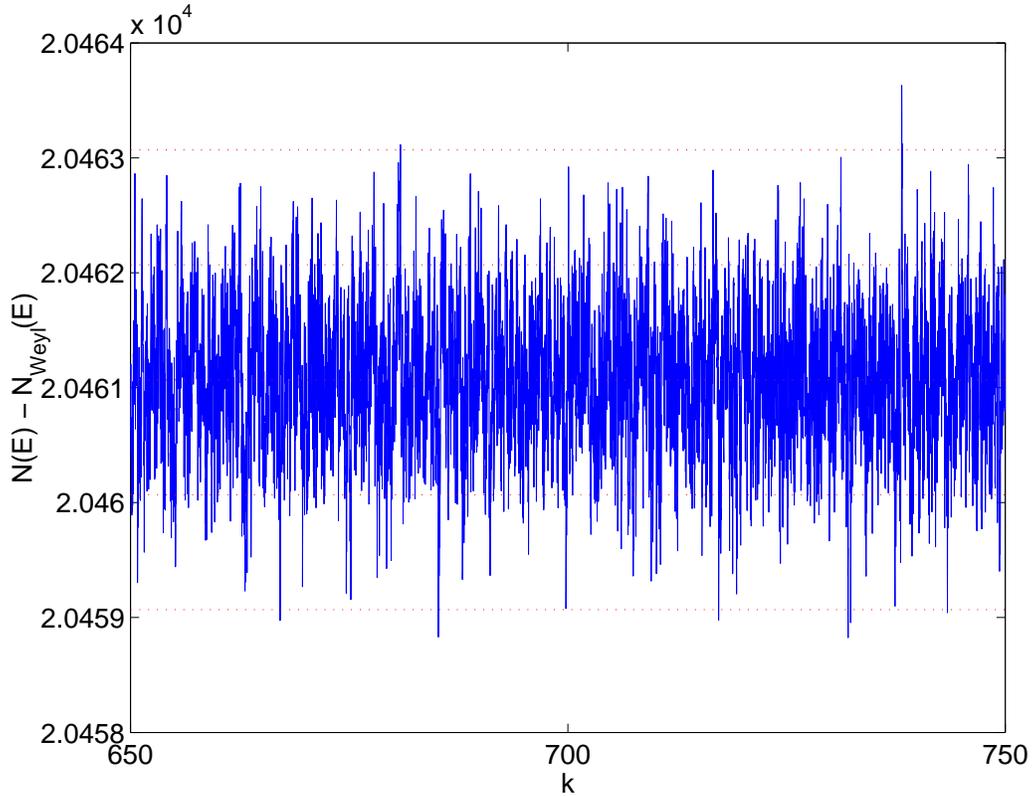}
\ec
\caption{Demonstration that none of the 6812 modes are missing in the
wavenumber window $k_n\in[650,750]$.
The level counting function (\ref{eq:ne}) is plotted
after the first two Weyl terms
$N_\tbox{Weyl}(E) = (\vol(\Omega)/4\pi) E - (|\Gamma|/4\pi) \sqrt{E}$
have been subtracted.
The horizontal axis shows wavenumber $k = E^{1/2}$.
Spectral rigidity ensures that a single missing or extra mode can
be detected~\cite{gutz}; this would be visible as a permanent jump of size 1
(the gap between the dotted horizontal lines). No such jump occurs.
\label{fig:weyl}
}
\end{figure}

\section{Computing quantum matrix elements}
\label{sec:num}

Our nonsymmetric
billiard $\Omega$ is shown in Fig.~\ref{fig:bil}a and defined in the
caption.
Its classical autocorrelation has already been studied~\cite{wlf}.
In App.~\ref{app:cl} we present the method used to compute $\cclaw$ to
an accuracy of a fraction of a percent (this is much less time-consuming
than the following quantum computations).
Note that since two walls are the coordinate axes
it is a desymmetrized version of a billiard of four times the area
formed by unfolding reflections in $x$ and $y$.
Desired eigenfunctions are then the subset of odd-odd symmetric
eigenfunctions of the unfolded billiard, a fact that enables
use of a smaller symmetrized basis set (hence higher eigenvalues),
and a reduction of the effective
boundary to $\gdes$, the part of $\Gamma$ which excludes the
symmetry lines (axes), accelerating the quantum calculation by a factor of
about 4~\cite{mythesis}.

We chose the test function $A\ofr$ to be the characteristic function
of the subdomain $\Omega_A$ shown in Fig.~\ref{fig:bil}b which
falls one side of the straight line $\partial \Omega_A \setminus \Gamma$.
Our choice of the shape of $\Omega_A$ was informed by the issue
of boundary effects raised by B\"{a}cker
\etal \cite{baecker}, the main point being that within a boundary layer
of order a wavelength, there are Gibbs-type
phenomena associated with spectral projections, and
by choosing a large angle of intersection of
the line with $\Gamma$ their contribution is minimized.
Our classical mean is $\Abar = \vol(\Omega_A)/\vol(\Omega) \approx 0.55000$.
Matrix elements $\Ael{n}{m} = \langle \phi_n,\phi_m \rangle_{\Omega_A}$
are computed using integrals of eigenfunctions in an efficent manner
described shortly.

Eigenfunctions and eigenvalues were found with the
`scaling method'~\cite{v+s,mythesis},
outlined further in App.~\ref{app:sca}. This is a little-known
basis approximation method, a variant of the Method of Particular
Solutions~\cite{mps,KS},
which uses collocation on $\gdes$ to extract all eigenvalues lying
in an energy window $E_n\in[E,E+L(E)]$, where $L(E)$ is $O(E^{1/2})$.
The equivalent wavenumber window is $O(1)$.
The spectrum in larger intervals can then be found by collecting
from sufficiently many windows.
We are certain that all modes and no duplicates have been
found in the desired intervals, as demonstrated by comparing
the counting function against Weyl's law in Fig.~\ref{fig:weyl}.
At energy $E$ (wavenumber $k$) the
required basis size $N = O(\nsc)$, the `semiclassical basis size', is
\be
	\nsc \;:=\; \frac{k|\gdes|}{\pi} ,
\label{eq:basisN}
\ee
where $|\gdes|$ indicates the desymmetrized perimeter.
Rather than evaluating $\phi_n$ at a set of points covering all of $\Omega$,
which would be very expensive, only
{\em basis coefficients} of $\phi_n$ are computed, from which $\phi_n\ofr$
can be later computed at any desired location.
Computational effort is $O(N^2) = O(E)$ per mode,
assuming $O(N^3)$ theoretical effort for dense matrix diagonalization
(in fact because of memeory limitations, for large $N$ it was not this
favorable).
Because of the simultaneous computation of many modes, the scaling method is
faster by
$O(N)$ than any other known method (see overview in~\cite{KS,mythesis})
such as boundary integral equations~\cite{baeckerbim}.
We needed $N \approx 3500$ at the largest wavenumber
reached ($k_n\approx 4000$, at $n\approx 7\times 10^5$)
for the billiard under study; in this
case the resulting efficiency gain is roughly a factor of a thousand!
Similar efficiency gains have been reported in other studies
of the Dirichlet eigenproblem at extremely high energy in
both 2D~\cite{v+s,scalinguse1,casati} and 3D~\cite{prosen3d}.
Only a couple of studies in billiards have computed eigenfunctions at greater
$n$, and they invariably involved shapes without corners
(for example~\cite{casati}). Note that in App.~\ref{app:sca} we
outline the basis set innovation that allows us to handle non-convex shapes
and (non-reentrant) corners effectively.
Despite its success the scaling method has not yet been analysed
in a rigorous fashion~\cite{scaling}.

Once a large set of eigenfunctions (in the form of their basis coefficients)
have been found, such as the example plotted in Fig.~\ref{fig:efunc},
matrix elements may be efficiently computed as we now show.
The following is proved in App.~\ref{app:id}.
\begin{lem}\label{lem:overlap2d}
Fix $E>0$ and let
$-\Delta u = Eu$ and $-\Delta v = E v$ hold in a Lipschitz domain
$\Omega_A\in\mathbb{R}^2$.
Let $\mbf{n}$ be the outwards unit normal vector at
boundary location $\mbf{r}\in\partial\Omega_A$
($\mbf{r}$ is measured relative to some fixed origin),
and $ds$ be surface measure. Then
\be
	\langle u,v \rangle_{\Omega_A}  = 
	\frac{1}{2E} \oint_{\partial\Omega_A}
	\!\! (\mbf{r}\cdot\mbf{n})(Euv - \nabla u \cdot \nabla v)
	+ (\mbf{r}\cdot\nabla u)(\mbf{n}\cdot\nabla v)
	+ (\mbf{r}\cdot\nabla v)(\mbf{n}\cdot\nabla u)
	\; ds. \nonumber
\ee
\end{lem}
This boundary integral identity, with the substitution $u=v=\phi_n$,
allows diagonal matrix
elements to be calculated using 1D rather than 2D numerical
integration. Since typically 10 quadrature points per wavelength
are needed for integration,
and our system is up to hundreds of wavelengths in size,
this is an enormous efficiency gain of order $O(N)$ or $10^3$.
Note that the boundary integrand is nonzero only on the line
$\partial\Omega_A\setminus\Gamma$.
Off-diagonal matrix elements are found via
the following identity which is a simple consequence of the Divergence Theorem.
\begin{lem}
Let $-\Delta u = E_u u$ and $-\Delta v = E_v v$ hold with $E_u\neq E_v$,
and other conditions as above, then
\be
	\langle u,v \rangle_{\Omega_A} \; = \; \frac{1}{E_u - E_v}
	\oint_{\partial\Omega_A} \!
	(u \mbf{n}\cdot\nabla v - v \mbf{n}\cdot\nabla u)
	\; ds. \nonumber
\ee
\end{lem}
Again we use $u=\phi_n$, $E_u=E_n$, $v=\phi_m$, $E_v=E_m$, and the
integrand is nonzero only on $\partial\Omega_A\setminus\Gamma$.

Thus 
values and first derivatives
of eigenfunctions on boundaries alone
are sufficient to evaluate all matrix elements.
Note the eigenfunctions need never be evaluated in the interior of $\Omega_A$.
For the boundary integrals, $O(N)$
quadrature points are needed, and at each point $O(N)$ basis evaluations
are needed to find $\phi_n$ and its gradient, giving
$O(N^2)$ effort per eigenfunction, the same effort required to find
modes by the scaling method.
The calculations reported in this work
took only a few CPU-days (1GHz Pentium III
equivalent, 1--2 GB RAM)
in total. The effort is divided roughly equally between
evaluating basis (Bessel)
functions and their gradients at the quadrature points,
and dense matrix diagonalization.

\begin{figure}[tbp]
\bc
\includegraphics[width=\textwidth]{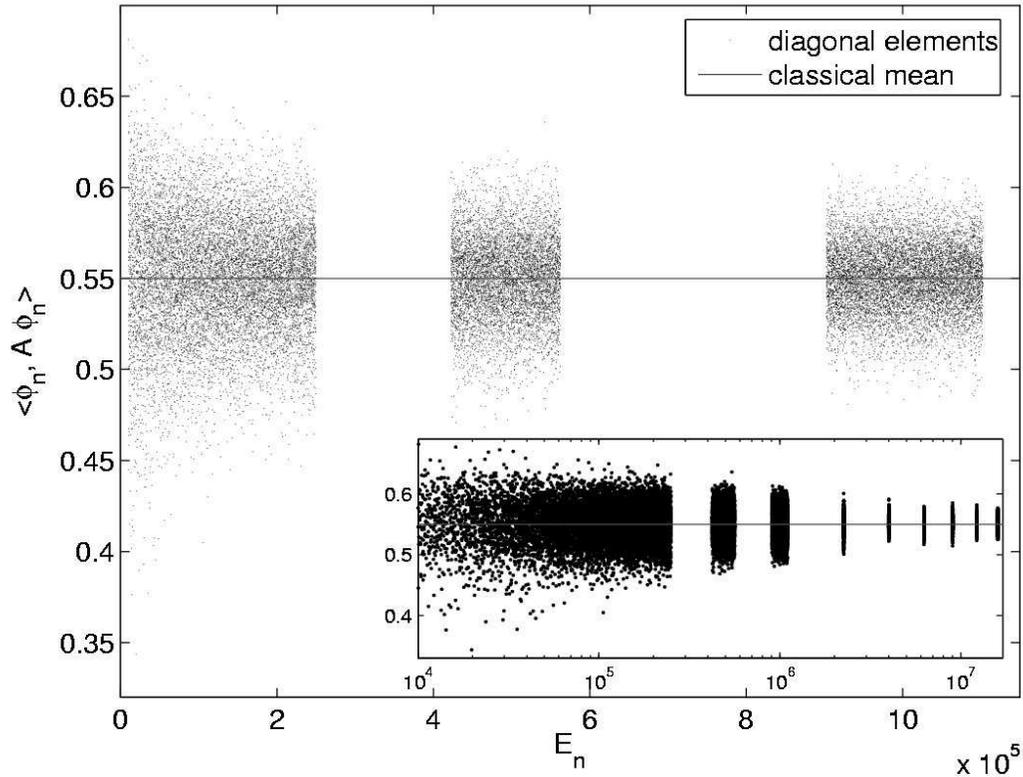}
\ec
\caption{Scatter plot of diagonal matrix elements $\Ael{n}{n}$ plotted against
energy eigenvalue $E_n$. The gaps are due to the fact that
only certain windows on the $E$ axis have been computed;
within each window all eigenvalues are found.
The windows shown in the main plot
correspond to wavenumbers $k_n\in[100,500]$, $k_n\in[650,750]$
and $k_n\in[950,1050]$,
giving a total of 28171 modes. 
The classical mean
$\Abar$ is shown as a horizontal line.
The inset shows the complete energy range
(including the 2718 higher modes not shown in the main plot)
on a log scale, with larger points to make extreme values evident. 
\label{fig:rawmc}
}
\end{figure}
\begin{figure}[tbp]
\bc
\includegraphics[width=\textwidth]{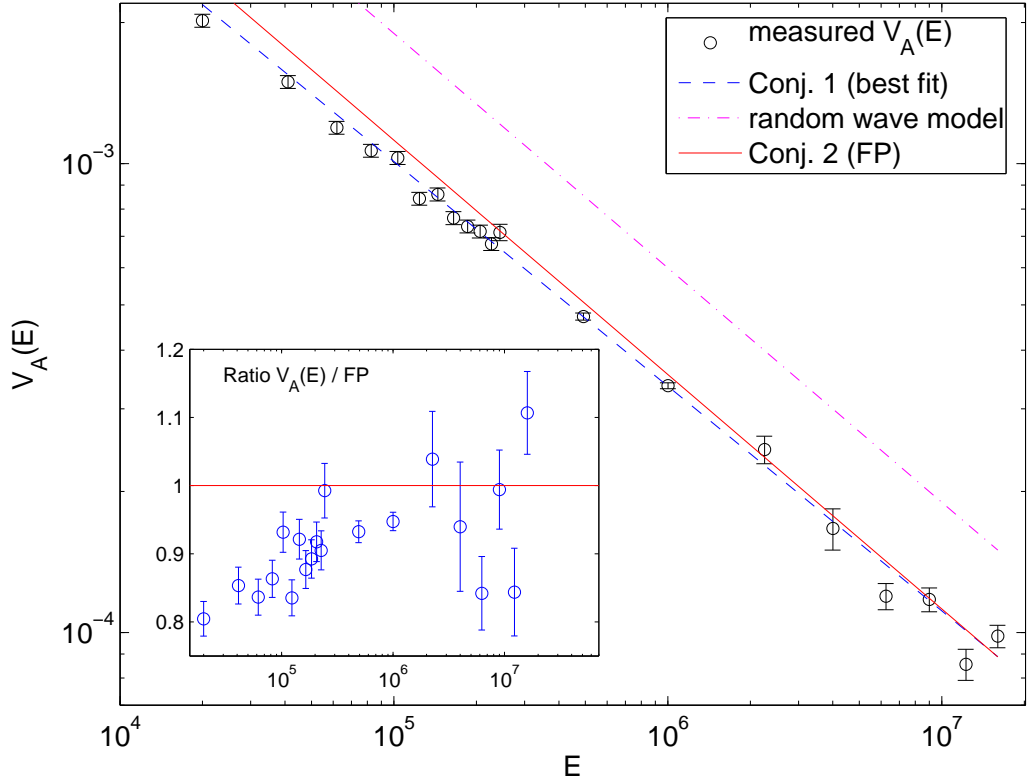}
\ec
\caption{Log-log plot of
diagonal variance $V_A(E)$ (with $\pm 1\sigma$ errorbars)
as a function of energy $E$.
The smallest errorbar is about 1.4\% (at $E\approx 10^6$).
For \cp\  the best-fit power law $(\gbf,\abf)$
is shown as the dashed line.
The random wave prediction (\ref{eq:rwg}) and (\ref{eq:rwa})
is the dash-dotted line.
\cd\  (FP) is shown as a solid line,
and the ratio of $V_A(E)$ to this prediction is shown in the inset.
\label{fig:fit}
}
\end{figure}

\begin{table}
\begin{tabular}{l|l|l|l|l|}
& Best-fit & Random wave & Feingold-Peres \\
& $\abfh$ & $\arw$  & $\afp$ \\
\hline
prefactor & $0.334\pm0.003$ & $0.5995 \pm 0.001$ & $0.3550\pm0.0004$\\
\hline
deviation from $\abfh$ & --- & $79\pm1\%$  & $6.5\pm0.9\%$ \\
\hline
\end{tabular}
\caption{
Comparison of best-fit prefactor $a$ in \cp,
assuming fixed power $\gamma = 1/2$, against the random wave prediction
(\ref{eq:rwa}), and the Feingold-Peres prefactor
$\afp := 2 \tilde{C}_A(0)/\vol(\Omega)$ of \cd.
\label{tbl:abf}
}
\end{table}

\section{Results and discussion}
\label{sec:res}

\subsection{Diagonal variance
(Conjectures~\ref{conj:power} and \ref{conj:diag})}
\label{sec:diag}

Fig.~\ref{fig:rawmc} shows a sample of raw diagonal matrix element data,
using only eigenvalues in certain intervals, up to $k\approx 10^3$
($E\approx 10^6$).
From this we
chose a sequence of $E$ values and computed $V_A(E)$ at each.
Computing matrix elements at high eigenvalues is very costly,
so it would be inefficient and impractical
numerically to grow the interval width as
$L(E) = cE^{1/2}$ for some constant $c$.
This would either involve ignoring, or breaking in to very short intervals
(which would introduce large relative fluctuations)
low eigenvalue data which
is cheap to collect, or requiring vast numbers of modes at high eigenvalue
which is too expensive.
Rather, we chose a convenient value of $L(E)$ at each $E$
that allowed a large number of modes $M$
to be summed at that $E$, while still allowing access to the highest $E$ values
possible ($E \approx 1.6\times10^7$).
Specifically, we split the lowest interval shown ($k_n\in[100,500]$) into
intervals containing about $10^3$ modes each, kept the two
other intervals shown intact (each containing an extremely large number of
order $10^4$), and at the higher eigenvalues
(see inset of Fig.~\ref{fig:rawmc}) chose
intervals containing 200--700 modes each.

One may ask how the $V_A(E)$ values obtained this way would differ from those
obtained using a strictly growing $L(E) = cE^{1/2}$.
To indicate the flutuations in $V_A(E)$ expected from summing over a
finite sample size $M$,
we included an {\em errorbar} of relative size $\sqrt{2/M}$.
For example $M=20000$ corresponds to 1\% errorbar (illustrating that
high statistical accuracy is computationally intensive). This model does assume
that deviations $|\Ael{n}{n} - \Abar|$ are roughly statistically
independent, an assumption motivated by RMT~\cite{RMT}.
We find (see inset of Fig.~\ref{fig:fit})
that observed fluctuations in $V_A(E)$ fit this assumption well, so that even
if correlations are present they do not affect our conclusions much.
Thus if longer sums were indeed computed using $L(E) = cE^{1/2}$,
it is likely that their values would lie within the errorbars shown.
The resulting local variance is presented in Fig.~\ref{fig:fit}.
A power-law dependence is immediately clear.
We fitted the power-law model of \cp, obtaining
a best-fit power
\be
	\gbf \;=\; 0.479 \pm 0.009,
\label{eq:gbf}
\ee
differing by only $4\pm 2\%$ from the random-wave model
and \cd\  value $\gamma=1/2$.
For this fit we used weighted least-squares, weighted using the
numbers of modes $M$ (equivalently, errorbars) for each interval.
In a standard maximum-likelihood framework~\cite{bayes} we
marginalized over $a$ to obtain the quoted errorbar in $\gamma$.
We also excluded a low-eigenvalue regime found to be non-asymptotic,
taking only data with $E>E_\tbox{min} = 1.6\times 10^5$.
(Our criterion here was that if any lower $E_\tbox{min}$ was used, the
fitted $\gbf$ was found to dependent on $E_\tbox{min}$).
This best-fit power-law is shown in Fig.~\ref{fig:fit};
the data are completely consistent with \cp.

The random wave model and \cd\  both involve the power
$\gamma=1/2$. Therefore to test their validity the power was held fixed at
this value while fitting only for the prefactor $a$, with results given in
Table~\ref{tbl:abf}.
Notice that the random wave model is a poor predictor;
this might be expected since this model takes no
account of the boundary conditions, yet
the support of the test function $A$ extends to the boundary and covers
a large fraction of the volume.
Intuitively speaking, images (boundary reflections) of $A$ are not being taken
into account, and evidently this is a large effect.

The prefactor (which is dominated by the large numbers of modes
at $E\approx 10^6$)
is overestimated by \cd\  by only
6.5\%, however this is statistically significant (a $7\sigma$ effect).
The systematic deviation is highlighted in the inset of
Fig.~\ref{fig:fit}, where it appears that the deviation decreases with
$E$. This is especially clear for $E\approx10^6$ or below, and
asymptotic agreement with \cd\ 
is not inconsistent with the large scatter of the highest six datapoints
(since they have larger errorbars).
If \cd\  is correct, our results show that convergence must be
alarmingly slow. This may explain why previous numerical billiard
studies~\cite{aurich,baecker} found various power-laws and prefactors
differing by up to 20\% from \cd, depending on choice of billiard
and $A$: they had failed to reach the asymptotic regime.
We achieve this with mode numbers 100 times higher than these studies.
We might model slow convergence to \cd\  by a correction, for example,
\be
 	V_A(E) \;\sim\; \frac{g\tilde{C}_A(0)}{\vol(\Omega)} E^{-1/2}
	\left(1 - b E^{-\beta} + o(E^{-\beta})\right) ,
\label{eq:fpcorr}
\ee
with $\beta$ sufficiently small, and $b$ sufficiently large.
Our current data do not allow a meaningful fit for $\beta$ and $b$,
however, they suggest $0.25 \le \beta \le 0.5$, and that $b$ is
several times greater than 1.
We note that in the physics literature periodic-orbit corrections to \cd\ 
derived by Wilkinson~\cite{wilk87} were later claimed by Prosen
not to contribute at this order~\cite{prosen}.
Clearly more numerical and theoretical work is needed on this issue of
slow convergence.

\begin{figure}[tbp]
\bc
\includegraphics[width=\textwidth]{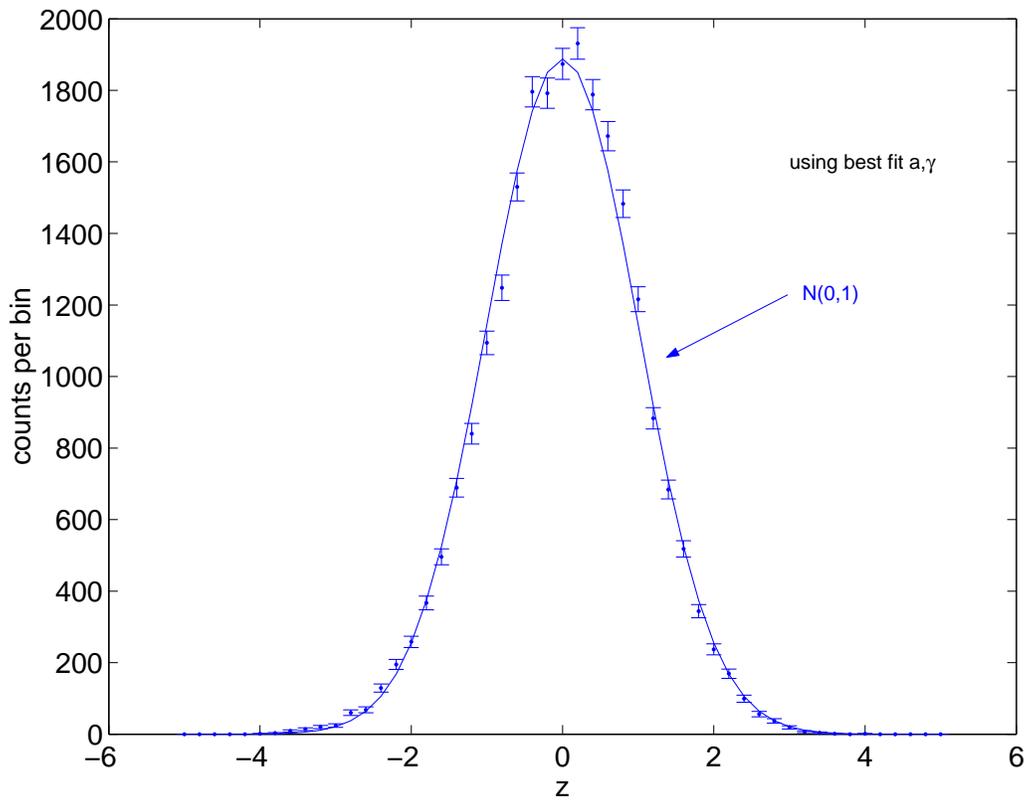}
\ec
\caption{Histogram of rescaled diagonal deviations
$z_n := (\Ael{n}{n} - \Abar)/V_A(E_n)^{1/2}$,
using the best-fit power law form for \cp,
compared against an appropriately-normalized
Gaussian distribution of unit variance.
\label{fig:hist}
}
\end{figure}

\subsection{Quantum Unique Ergodicity (\cq)}
\label{sec:que}

We now examine individual diagonal matrix elements.
The inset of Fig.~\ref{fig:rawmc} enables extreme values to be seen:
it is clear that there are no anomalous extreme values
which fall outside of a
distribution which is condensing to the classical mean.
Both the maximum 0.6811 and minimum 0.3437 of $\Ael{n}{n}$ occur at
$E_n < 2\times 10^4$, visible at the far left side.
This is strong evidence for QUE in this system.
Since about 30000 modes are tested, the density of any excluded sequence
can be given an approximate upper bound of $3 \times 10^{-5}$.
We note that it is possible (althouth unlikely) that
by an unfortunate choice of $A$,
non-uniform
or scarred modes occur which do {\em not} have
anomalous $\Ael{n}{n}$ values. The only way to eliminate this
possibility would be to repeat the experiment with a selection of different $A$
functions.

What is the distribution that the deviations $\Ael{n}{n} - \Abar$
follow? We have rescaled these deviations according to the
best-fit form of the variance in \cp, and histogram the results in
Fig.~\ref{fig:hist}.
The distribution is consisitent with a gaussian, with an excellent
quality of fit. (The slightly fatter tail on the low side is entirely
due to low-lying modes; as can be seen in Fig.~\ref{fig:rawmc} these
have a skewed distribution).

\begin{figure}[tbp]
\bc
\includegraphics[width=\textwidth]{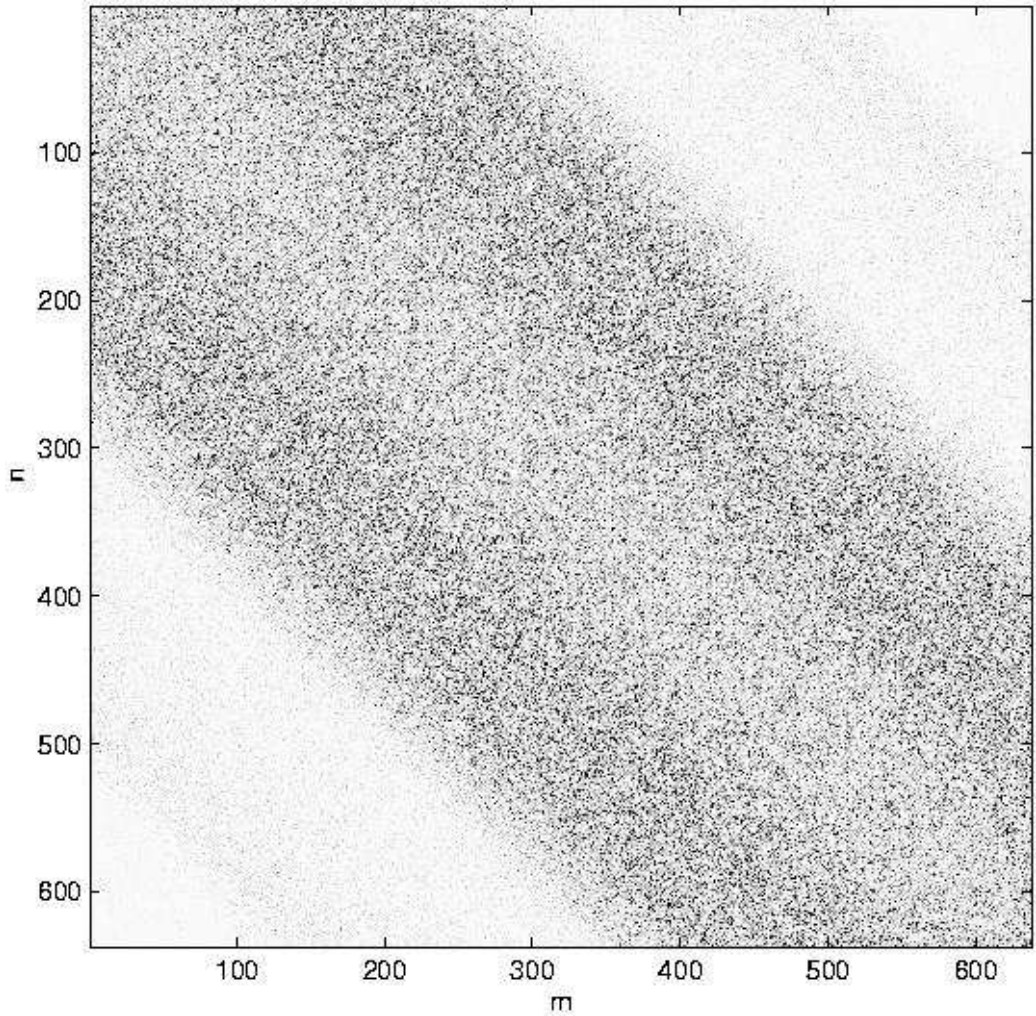}   
\ec
\caption{Density plot of block of squared matrix elements
$|\Ael{n}{m} - \delta_{nm}\Abar|^2$ for
the 637 modes lying in $k\in[650,660]$.
The range white to black indicates zero to $1.7\times10^{-3}$.
Individual elements appear uncorrelated, the only visible structure
being the intensity varying with spectral measure (band profile).
\label{fig:mat}
}
\end{figure}
\begin{figure}[tbp]
\bc
\includegraphics[width=\textwidth]{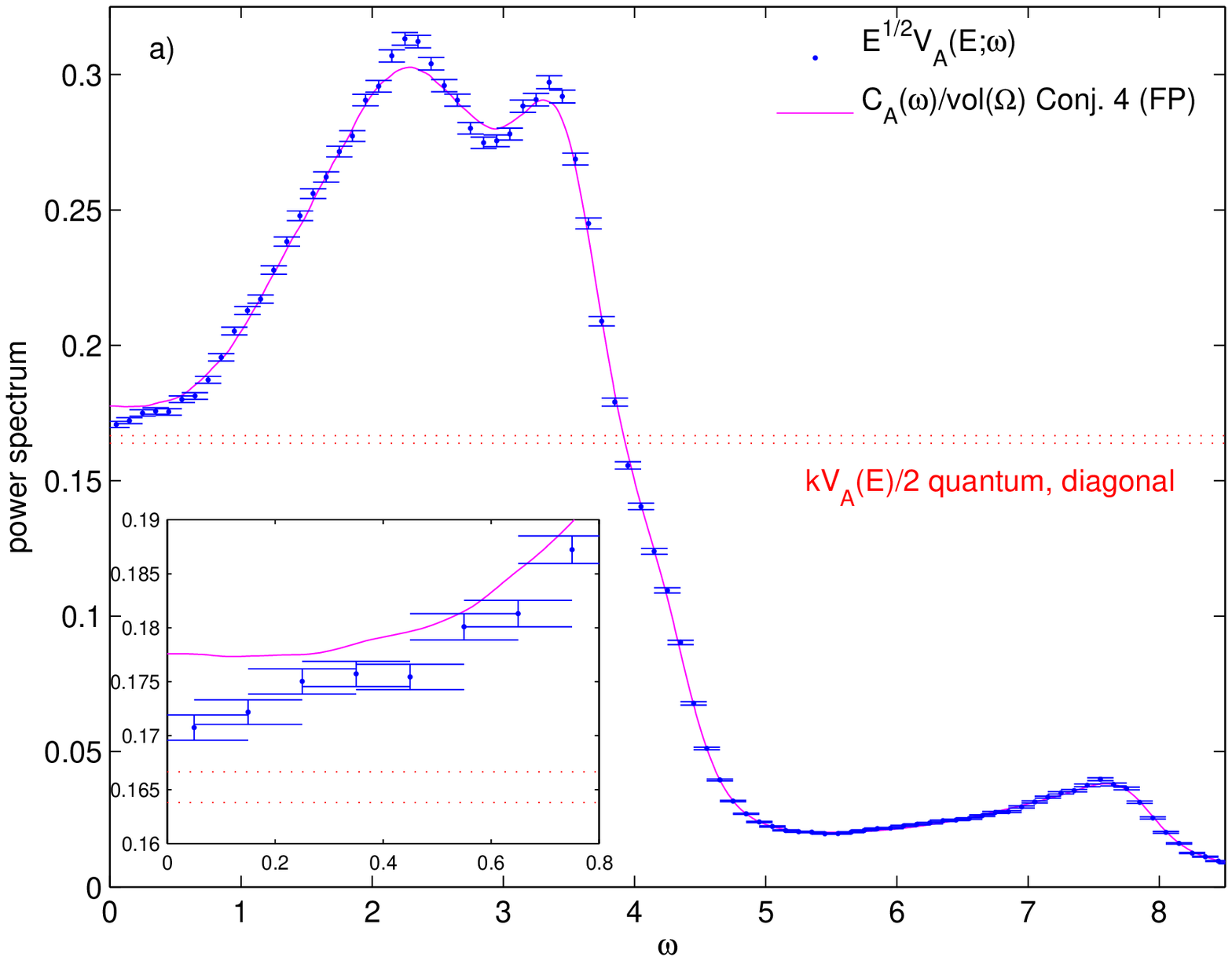}
\includegraphics[width=\textwidth]{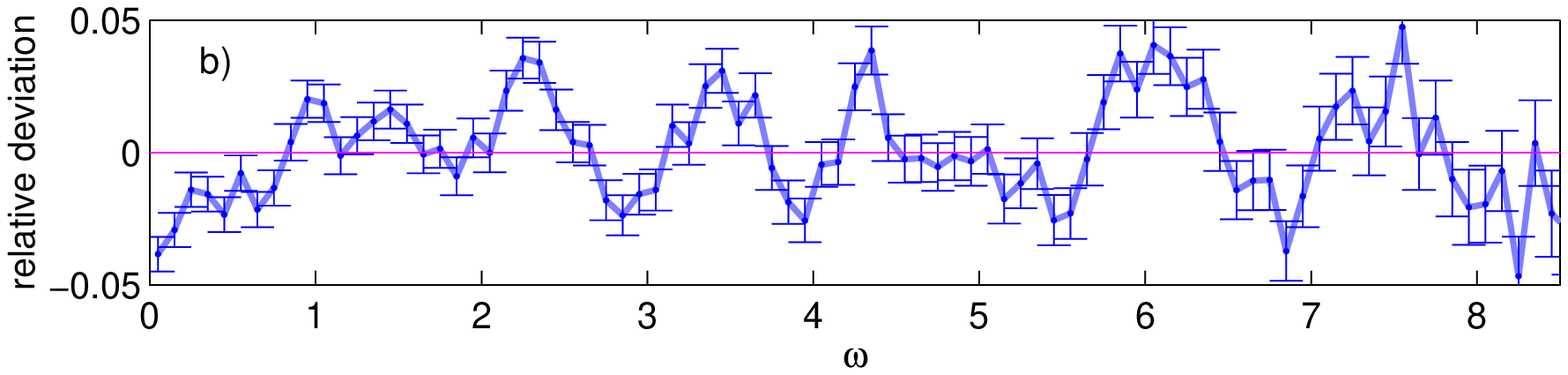}
\ec
\caption{a) Off-diagonal quantum variance $V_A(E;\omega)$ (errorbars)
at a single $E\approx4.9\times 10^5$,
as a function of distance $\omega$ (wavenumber units) from the diagonal,
compared against \co\ estimated using the classical power spectrum
(solid line).
Errorbars are 0.7\% for quantum (near the diagonal),
0.2\% for classical.
Also half the diagonal variance is shown
as two horizontal dotted lines indicating $\pm1$ standard deviation.
Inset is a zoom on the $\omega\to0$ region.
b) Relative deviation (difference from 1 of ratio)
between quantum variance and \co.
\label{fig:offdiag}
}
\end{figure}

\subsection{Off-diagonal variance (\co)}
\label{sec:offdiag}

We decided, for reasons of numerical practicality,
to test \co\ over a range of $\omega$, but
at a single (large) value of $E$. This was performed using the
single sequence of 6812 eigenfunctions
with $k_n\in[650,750]$, from which about 2.3 million individual
off-diagonal elements $\Ael{n}{m}$ were calculated, namely those in the block
$n,m: k_n, k_m\in[650,660]$, the block
$n,m: k_n, k_m\in[660,670]$, etc, with 10 blocks in total.
The first such matrix block is plotted in Fig.~\ref{fig:mat}; note the
strong diagonal band structure (`band profile'~\cite{doronfrc}).
The mean off-diagonal element variances lying in
successive $\omega$-intervals of width 0.1 were then collected.
Thus we are testing \co\ with a window of $L(E) \approx 1.4\times 10^5$
and $\epsilon(E) = 0.05$.
Our extremely large choice of $L(E)$ allowed statistical fluctuations
to be minimized (errorbars were estimated as in Section~\ref{sec:diag},
and are generally less than 1\%).
We believe this is the most accurate test of the
conjecture ever performed.

The resulting band profile $V_A(E;\omega)$ is compared against \co\
in Fig.~\ref{fig:offdiag}.
The agreement is excellent, with generally less than 3\% discrepancy.
Let us emphasise that there are no fitted parameters.
There appears to be statistically significant deviations:
the peaks and valleys (points of highest curvature, including $\omega=0$)
are exaggerated more in the quantum variance than the classical.
Care has been taken to ensure that this was not due to
numerical errors in $\cclaw$ (it was smoothed only on a much finer scale,
see Appendix~\ref{app:cl}).
As with the diagonal variance, this may reflect slow convergence.
The diagonal quantum variance, divided by $g=2$, is also shown. If step (ii)
of Section~\ref{sec:fp} applied exactly then this would coincide
with $V_A(E,0)$, however a $3\pm 1\%$ difference is found.
It is unlikely
but possible that this is merely a statistical fluctuation (a null result).
Thus, at this energy, the diagonal variance prefactor discrepancy
of 7\% seems to result from the addition of
two roughly equal effects: step (i) has about 4\% discrepancy, and step
(ii) about 3\%.
Our data does not rule out the possibility that $g$ is closer to 1.85 than
to 2, which would imply slight positive correlations between
neighboring eigenfunctions.

In terms of individual elements, we find no anomalously large values.
This strongly supports `off-diagonal QUE': the vanishing of
every single off-diagonal element, a stronger result than~\cite{zeloffdiag}.
A preliminary examination suggests uncorrelated
gaussian distribution of elements, with variance given
by $\omega$-distance from the diagonal, but we
postpone analysis for future work.

\subsection{Discussion on existence of scars}
\label{sec:scar}

In the physics community the existence of
scars is well-known, as are theoretical (non-rigorous) models.
Heller~\cite{hel84,hellerhouches} put forward
a semiclassical explanation based on enhanced short-time
return probability for wavepackets launched along the least unstable
periodic orbits (UPOs),
which has been elaborated~\cite{bogscar,berryscar,agam,lev}.
Although the meaning of `scar' varied historically, it is now taken
to mean {\em any} deviation from the random wave
prediction of eigenfunction intensity near a UPO~\cite{lev}.
Scar `strength' depends
on what test function you use to measure it~\cite{scarmometer}:
in physics this test function is commonly {\em not} held fixed as the
limit $E\to\infty$ is taken,
rather it is chosen to collapse microlocally onto a UPO with a
coordinate-space width $\sim E^{-1/4}$.
By this measure, scar strength is believed
not to die out in the semiclassical limit~\cite{lev}.
Typical scar intensities ($|\phi_n|^2$ along the UPO) do not decay,
but their width,
hence the associated probability mass, vanishes
as $O(E^{-1/4})$.

However, in the mathematics community
questions of uniformity of eigenfunctions and QUE are presented in
the form of weak limits, that is, limits of matrix elements of
fixed, $0^{th}$-order pseudo-differential operators.
Persistence of scarring is taken to mean existence of a subsequence
with $|\Ael{n}{n} - \Abar| = O(1)$.
Thus we might distinguish a {\em physicist's scar}, where a
probability mass vanishing as $O(E^{-1/4})$ is associated with the UPO,
from a {\em mathematician's scar} (or `strong scar') which carries
$O(1)$ probability mass (as in~\cite{RS}).

Our results (Section~\ref{sec:diag}) suggest strong scars do not persist
asymptotically, but is consistent with the persistence of
physicist's scars, in fact giving the same power-law
$|\Ael{n}{n} - \Abar| = O(E^{-1/4})$ expected from scar width.
Heller's numerical demonstrations of apparently strong scarring
were done at mode numbers $n\approx2\times 10^3$,
which, in light of our work, is well below the asymptotic regime.
It is now believed by physicists that in 2D Anosov billiards
strong scarring does not persist~\cite{agam,lev,li}, but
there still exist controversies about the width of scars~\cite{li},
and in related quantum models
the mechanism of scarring is an active research area~\cite{schanz}.
Mathematically, the issue remains open.

\section{Conclusions}
\label{sec:conc}

We have studied a generic Euclidean Dirichlet billiard whose
classical dynamics is Anosov, and studied quantum
ergodicity of both diagonal and off-diagonal matrix elements.
By accessing very high eigenvalues
(100 times higher than previous studies) using the scaling method
and boundary integral formulae for $\Ael{n}{m}$ in the case of
piecewise-constant test function $A$,
we believe we have reached the asymptotic regime for first time.
We also have unprecedented statistical accuracy due to the large number
of modes computed.
A summary of evidence found for the four conjectures
from the Introduction is as follows:
\bi
\item \cp~: Diagonal variance shows
excellent agreement, with power-law $\gamma = 0.479$.

\item \cd~: Diagonal variance is consistent with the Feingold-Peres
prediction,
but with quite slow convergence (for instance, a 7\% overestimate remains
at $E=10^6$).
This contrasts a random-wave model, which overestimates
the fitted prefactor by 80\%.

\item \cq~: Compelling evidence for QUE in this system
(density of exceptional subsequence $<3\times 10^{-5}$).

\item \co~: Excellent agreement for off-diagonal variance (of order $3\%$
discrepancies at $E=5\times10^5$), and evidence for off-diagonal QUE.
\ei

Performing this large-scale study required limiting ourselves to
one billiard $\Omega$ and one operator $A$. In order to complete the picture,
the discrepancy between the predictions of
Conjectures~\ref{conj:diag} and \ref{conj:offd}
and numerically-measured variances should be
studied as a function of $\Omega$ and $A$.

\section*{Acknowledgements}
This work was inspired by questions of Peter Sarnak, with whom the
author has had enlightening and stimulating interactions.
We are also delighted to have benefitted from discussions
with Steve Zelditch, Percy Deift,
Fanghua Lin, Eduardo Vergini, Doron Cohen, Eric Heller,
Kevin Lin, and the detailed and helpful comments of the anonymous reviewer.
While the bulk of this work was performed, the
author was supported by the Courant Institute at New York University.
The author is now supported by NSF grant DMS-0507614.

\appendix

\section{Classical power spectrum}
\label{app:cl}

We use standard techniques~\cite{stoc} to estimate $\cclaw$.
For a particular trajectory, launched with certain initial location in
phase space, $A(t)$ is a noisy function (stochastic stationary process).
We define its windowed Fourier transform as
\be
	\tilde{A}(\omega) \; := \; \int_0^T \! A(t) e^{i\omega t} dt,
\label{eq:aw}
\ee
where the window is a `top-hat' function from $0$ to $T$.
Using with (\ref{eq:ctraj}) and (\ref{eq:cw}), and taking care with order of
limits, we have the Wiener-Khinchin Theorem,
\be
	\cclaw \; = \; \lim_{T\rightarrow\infty} \frac{1}{2\pi T}\,
	\tilde{A}^*(\omega) \tilde{A}(\omega).
\label{eq:wiener}
\ee
For this single trajectory, $\tilde{A}(\omega)$
is a rapidly-fluctuating random function of $\omega$, with zero mean
(for $\omega\neq0$), variance given by $2\pi T \cclaw$,
and correlation length in $\omega$ of order $2\pi/T$. (As $T\rightarrow\infty$,
the $\omega$-correlation becomes a delta-function).
Thus (\ref{eq:wiener}) converges
only in the weak sense, that is, when smoothed
in $\omega$ by a finite width test function.

A given trajectory is found by solving the particle collisions
with the straight and circular sections of $\Gamma$, and $A(t)$ is
sampled at intervals $\Delta t = 0.02$ along the trajectory
(recall we assume the particle has unit speed).
Then $\tilde{A}(\omega)$ is estimated using the
Discrete Fourier Transform (implemented by an FFT library)
of this sequence of samples, giving
samples of the spectrum at $\omega$ values
separated by $\Delta \omega = 2\pi/T$.
The correlation in $\omega$ is such that each sample is (nearly) independent.
$\Delta t$ was chosen sufficiently short that
aliasing (reflection of high-frequency components into apparently low
frequencies) was insignificant.
A trajectory length $T=10^4$ (about $1.8 \times 10^4$ collisions)
was used.
The finiteness of $T$ causes relative errors of order $t_\tbox{corr}/T$,
where $t_\tbox{corr}\approx 2$ (for our domain) is the timescale for
exponential (since the billiard is Anosov) decay of correlations.
Thus more sophisticated window functions are not needed.

Given $\tilde{A}(\omega)$ we use (\ref{eq:wiener}), with
the $T$ given above, to estimate
$\cclaw$. We smooth in $\omega$ by a Gaussian of
width $\osm = 0.03$. This width is chosen to be
as large as possible to average the largest number of independent samples
from the neighborhood of each $\omega$, but
small enough to cause
negligible convolution of the sharpest features of $\cclaw$.

Finally, in order to reduce further the random fluctuations in the
estimate, $n_r=6000$ independent trajectory realizations with
random initial phase space locations were averaged.
An estimate for the resultant relative error $\epsilon$ in $\cclaw$
can be made by counting the number $N$ of independent random
samples which get averaged, and using the fact that
the variance of the square of a Gaussian zero-mean random
variable (\ie
$\chi^2$ distribution with 1 degree of freedom) is twice the mean.
This gives
\be
	\epsilon \; = \; \left(\frac{2}{N}\right)^{1/2}
	\;\approx \; \left(\frac{2\pi}{n_r \osm T}\right)^{1/2},
\ee
which numerically has been found to be a conservative estimate.
In our case $\epsilon \approx 2\times 10^{-3}$, that is, about 0.2\%
error.

The zero-frequency limit $\ccla(0)$ is found using the smoothed 
$\cclaw$ graph at $\omega=0$, and therefore is an average of frequencies
within $O(\osm)$ of zero. This is justified because
correlation decay (weak mixing) causes all moments of $C_A(\tau)$
to be finite, hence there is no singularity in $\cclaw$ at $\omega=0$ (it
can be expanded in an even Taylor series about
$\omega=0$ with finite coefficients).


\section{Scaling method for the Dirichlet eigenproblem}
\label{app:sca}

The scaling method for the solution of the Dirichlet eigenproblem in
star-shaped domains
was invented by Vergini and Saraceno~\cite{v+s}, and
considering its great efficiency 
it has received remarkably little attention. Here we give only an outline.

The method relies on the remarkable fact that
the normal derivatives of eigenfunctions lying close in energy
are `quasi-orthogonal' (nearly orthogonal)
on the boundary, with respect to the boundary weight function
$r_n:=\mbf{r}\cdot\mbf{n}$.
\begin{lem}\label{lem:quasi}
Let $\Omega\in\mathbb{R}^d$, $d\ge2$ be a Lipshitz domain with boundary
$\Gamma$ and Dirichlet spectral data $\{E_j\}$, $\{\phi_j\}$.
Let $\mbf{n}$ and $\mbf{r}$ be defined as in Lemma~\ref{lem:overlap2d},
with $r_n:=\mbf{r}\cdot\mbf{n}$. Then, for all $i\ge 1$, $j\ge 1$,
\be
	Q_{ij} := \oint_{\Gamma} \!
	r_n
	(\mbf{n}\cdot \nabla \phi_i)(\mbf{n}\cdot \nabla \phi_j)\,ds
	\; = \; 2E_i \delta_{ij} +
	\frac{(E_i-E_j)^2}{4}\langle\phi_i, r^2\phi_j\rangle_{\Omega},
\label{eq:quasi}
\ee
\end{lem}
This is proved in App.~\ref{app:id}.
A corollary is that, since $r^2:=\mbf{r}\cdot\mbf{r}$
is a bounded operator on the domain, off-diagonal
elements of $Q$ must vanish quadratically as one approaches
the diagonal. Thus the matrix with elements
$Q_{ij}/2E_i$ approximates the identity matrix, when restricted to
an energy window $E_i, E_j \in [E-\ep_0, E+\ep_0]$, if the window
size remains relatively narrow $\ep_0 = o(E^{1/2})$.

We choose a `center' wavenumber $k = E^{1/2}$, near which we are
interested in extracting eigenfunctions, and relative to
which the wavenumber shift of mode $i$ is
$\omega_i(k) := k-k_i$. Consider an
eigenfunction $\phi_i$ for which $\omega_i < 0$ and $|\omega_i| \ll O(1)$.
We create a version
spatially rescaled (dilated about the origin)
by an amount needed to bring its wavenumber to $k$,
that is, $\chi_i\ofkr  :=  \phi_i(k\mbf{r}/k_i)$.
We call this function $k$-rescaled.
Thus we have $-\Delta \chi_i = E \chi_i$ everywhere inside
$\Omega$, with $\chi_i\ofkr = 0$ on the rescaled boundary (\ie for all
$k\mbf{r}/k_i \in \Gamma$).
The rescaled eigenfunction can be Taylor expanded in $\omega_i$,
\bea
	\chi_i\ofkr & = &
	\phi_i\left(\mbf{r} + \frac{\omega_i}{k_i}\mbf{r}\right)
	\; = \; \phi_i\ofr + \frac{\omega_i}{k_i}\mbf{r}\cdot\nabla\phi_i
	+ O(\omega^2) \nonumber \\
	& = & \frac{\omega_i}{k_i}r_n\mbf{n}\cdot\nabla\phi_i + O(\omega^2)
	\qquad \mbox{for } \mbf{r}\in\Gamma ,
\label{eq:exp}
\eea
where Dirichlet boundary conditions were applied.
We construct a basis of $N$ functions $\xi_l\ofkr$, satisfying
$-\Delta \xi_l = E \xi_l$ inside $\Omega$, no particular
boundary conditions on $\Gamma$, and non-orthogonal over $\Omega$.
We assume they approximately span the linear space
in which rescaled eigenfunctions live, so that
\be
	\chi_i\ofkr \; = \; \sum_{l=1}^N X_{li} \,\xi_l\ofkr
	 \; + \;\epsilon_i\ofr	\qquad \mbox{for all $i$ of interest},
\label{eq:basis}
\ee
where the error $\epsilon_i$ can be made negligibly small for some $N$.
In practise $N$ need exceed $\nsc$ defined in (\ref{eq:basisN})
by only a small factor (2 or less).
Our goal is then to solve for a shift $\omega_i$ and
the corresponding $i^{th}$ column of the coefficient matrix $X$.
We can do this by simultaneous diagonalization of quadratic forms.
We define two symmetric bilinear forms on the boundary,
\bea
	f(u,v) & := & \oint_\Gamma \frac{1}{r_n} u v\, \ds, \\
\label{eq:fuv}
	g(u,v) & := & \frac{1}{k}\oint_\Gamma \frac{1}{r_n}
	(u\mbf{r}\cdot\nabla v + v\mbf{r}\cdot\nabla u)\, \ds.
\label{eq:guv}
\eea
Note that defining these forms
brings the extra requirement that the domain
be strictly star-shaped about the origin
($r_n > 0$), which from now on we assume.
In the rescaled eigenbasis $f$ is,
via (\ref{eq:exp}) and (\ref{eq:quasi})
\be
	f(\chi_i, \chi_j) \; = \; \frac{\omega_i\omega_j}{k_i k_j}
	Q_{ij} + O(\omega^3)
	\; = \; 2\omega_i^2 \delta_{ij} + O(\omega^3),
\label{eq:f}
\ee
a matrix which is close to
diagonal, because of the closeness of $Q$ to the identity.
In the same basis, recognizing that for $k$-rescaled functions
$g$ is equivalent to $df/dk$, the
derivative of (\ref{eq:fuv}) with respect
to the center wavenumber, and using $d\omega_i/dk = 1$, we have
\be
	g(\chi_i, \chi_j) \; = \; \frac{\omega_i+\omega_j}{k_i k_j}
	Q_{ij} + O(\omega^2)
	\; = \; 4\omega_i \delta_{ij} + O(\omega^2),
\label{eq:g}
\ee
so $g$ is also close to diagonal.
Thus the set $\{\chi_i\}$ with small $|\omega_i|$
approximately diagonalizes both bilinear forms, with the approximation
error growing as a power of $|\omega_i|$.
As we explain below, in practise the converse applies,
that is, by simultaneously diagonalizing $f$ and $g$ we can extract
the set of eigenfunctions $\{\chi_i\}$ with smallest $|\omega_i|$.
Therefore, loosely speaking, when the boundary
weight function $1/r_n$ is used, domain
eigenfunctions
are given by the simultaneous eigenfunctions of the (squared) boundary norm
and its $k$-derivative.

We perform the diagonalization in the basis (\ref{eq:basis}).
That is, matrices $F_{lm}:=f(\xi_l,\xi_m)$ and $G_{lm}:=g(\xi_l,\xi_m)$,
with $l, m = 1\cdots N$,
are filled. This requires basis and
first derivative evaluations on the boundary.
It is an elementary fact that given a positive matrix $F$ and a symmetric
matrix $G$ there exists
a square matrix $Y$ and
a diagonal matrix $D := \mbox{diag}\{\mu_i\}$ which satisfy
$Y^T F Y = I$ and $Y^T G Y = D$.
The matrices $Y$ and $D$ can be found by standard numerical
diagonalization algorithms in $O(N^3)$ time.
If (\ref{eq:f}) and  (\ref{eq:g}) held without error terms, and
the number of modes $i$ for which they held
were equal to (or exceeded) the basis size $N$, then we would be able
directly to equate the columns of $Y$ with the desired columns of $X$
(barring permutations).
In this case $\omega_i  = 2/\mu_i$ would also hold, from which the desired
wavenumbers $k_i$ follow.
However, using Weyl's law and (\ref{eq:basisN}) is follows that
such a large number of modes requires that the largest $|\omega_i|$ is
of order unity, in which
case errors in (\ref{eq:f}) and (\ref{eq:g}) would become unacceptable.
It is an empirical observation found through numerical study
that in fact columns of $Y$ corresponding to the {\em largest magnitude}
generalized eigenvalues
$\mu_i$ (and therefore the smallest shifts $|\omega_i|$)
do accurately match columns of $X$.
Thus perturbations by other vectors in the span of basis functions are small.
Further discussion is postponed to a future publication~\cite{scaling}.

We mention a couple of other implementation details.
Because the generalized eigenproblem turns out to be singular it is
truncated to its non-singular part~\cite{v+s,mythesis}.
If columns of
$Y$ are normalized such that $Y^T F Y = I$ holds then the resulting
eigenfunctions can be normalized over $\Omega$ by dividing
the $i^{th}$ column of $Y$ by $\sqrt{2}\omega_i$.
Depending on the choice of basis, spurious solutions can result; they
are easily identified because their norm over $\Omega$, computed by the
following Rellich identity,
is not close to 1.
\begin{lem}[Rellich]
With the definitions of Lemma~\ref{lem:quasi}, for all $j\ge 1$,
\be
        \frac{1}{2E_j} \oint_{\Gamma}\!
	r_n (\mbf{n}\cdot\nabla \phi_j)^2 \, ds \; = \; 1
\label{eq:bw84uu}
\ee\label{lem:rel}
\end{lem}
This identity is a special case of Lemma~\ref{lem:overlap2d} found by
substituting $\Omega_A=\Omega$, $u=v=\phi_j$, and $E=E_j$,
and recognising $\mbf{r}\cdot\nabla \phi_j = r_n \mbf{n}\cdot\nabla\phi_j$.
The maximum $|\omega_i|$ in which levels of useful accuracy
are found is of order $0.2/R$
where $R$ is the largest radius of the domain.
The lack of missing modes obtained with this method is illustrated
by Fig.~\ref{fig:weyl}.
There are several implementation issues and improvements that
we do not have space to discuss here~\cite{v+s,mythesis,scaling}.

A word about the basis set choice $\{\xi\ofkr\}$, $1\le i\le N$, is needed.
Until now plane waves (including evanescent plane waves~\cite{v+s})
or regular Bessel functions~\cite{casati} have been used.
These fail for non-convex domain shapes, or those with corners,
thus to tackle the domain in this study a basis of
irregular Bessel (\ie Neumann) functions,
placed at equal intervals along a curve $\Gamma^+$
exterior to $\Omega$, was developed by the author.
$\Gamma^+$ is defined as the set of points whose nearest distance to $\Gamma$
is $D$, with $kD = 7$ (roughly one wavelength distant).
This was found to handle (non-reentrant) corners successfully. 
It performs extremely well for all shapes that have been attempted so far.
The basis size $N$ is about $1.5 \nsc$ (see (\ref{eq:basisN})),
thus, depending on required
accuracy, about $N/20$ useful modes are found per dense
matrix diagonalization ($O(N^3)$ effort).
This is $O(N)$ faster than other boundary
methods; we remind the reader that $N$ is larger than $10^3$ in our work.

\section{Identities involving eigenfunctions of the Laplacian}
\label{app:id}

Let there be constants $E_u>0$, $E_v>0$, and
let $-\Delta u = E_u u$ and $-\Delta v = E_v v$
hold in a Lipschitz domain
$\Omega_A\in\mathbb{R}^d$, for some general dimension $d\ge2$.
The following expressions for the divergence of certain vector fields
result from elementary calculus.
By $\nabla\nabla u$ we mean the
second derivative tensor (dyad).
\bea
\nabla \cdot (v\nabla u) &=& -E_u uv + \nabla u \cdot \nabla v
\label{eq:12}
\\
\nabla \cdot (\mbf{r}uv) &=&
duv + u\mbf{r}\cdot\nabla v + v\mbf{r}\cdot\nabla u
\label{eq:3}
\\
\nabla\cdot(\mbf{r}\nabla u \cdot \nabla v) &=& d\nabla u \cdot \nabla v
+\nabla u\cdot \nabla\nabla v \cdot \mbf{r}
+\nabla v\cdot \nabla\nabla u \cdot \mbf{r}
\label{eq:4}
\\
\nabla\cdot[(\mbf{r}\cdot\nabla u) \nabla v] &=&
\nabla u \cdot \nabla v - E_v v\mbf{r}\cdot\nabla u
+\nabla v\cdot \nabla\nabla u \cdot \mbf{r}
\label{eq:56}
\\
\nabla\cdot(r^2 v \nabla u) &=&
2 v \mbf{r}\cdot\nabla u - E_u r^2 uv + r^2 \nabla u \cdot \nabla v
\label{eq:78}
\eea
First we prove Lemma~\ref{lem:overlap2d} (also see App.~H of \cite{mythesis}).
Consider the following four equations:
(\ref{eq:3}), (\ref{eq:4}), (\ref{eq:56}), and its counterpart obtained by
swapping $u$ and $v$.
Integrating each of these over $\Omega_A$, then applying
the Divergence Theorem, gives four expressions for
surface integrals in terms of domain integrals.
Substituting these into the
right-hand side of the expression in Lemma~\ref{lem:overlap2d}, and
setting $E_u=E_v=E$,
gives, after cancellation, the left-hand side.

A similar but more complicated
technique proves Lemma~\ref{lem:quasi}.
First we consider modes $i,j$ which are non-degenerate, that is, $E_i\neq E_j$.
Eight equations are needed: the four mentioned
above, then (\ref{eq:12}) and (\ref{eq:78}) and their counterparts swapping
$u$ and $v$.
Each should be integrated over $\Omega$ and the Divergence Theorem applied.
The following identity may then be verified
by substitution of the eight resulting equations into its right-hand side.
Using the
abbreviations $\varepsilon:=E_u-E_v$, ${\cal E} :=E_u+E_v$, $u_n = \mbf{n}\cdot
\nabla u$, $u_r = \mbf{r}\cdot\nabla u$, the identity to be checked is,
\begin{multline}
        \frac{\varepsilon^2}{4}\int_\Omega r^2 uv\, d\mbf{r}
        \; = \;
        \oint_{\Gamma}\!
        \frac{d-2}{2}(u_n v+v_n u)
        + \left(\frac{\cal E}{\varepsilon} -
	\frac{\varepsilon}{4}r^2\right)(u_n v-v_n u)\\
        + r_n\left(\frac{\cal E}{2} uv - \nabla u \cdot \nabla v\right)
        + u_r v_n + v_r u_n \, ds.
\label{eq:diff_gen}
\end{multline}
The substitutions $u=\phi_i$, $v=\phi_j$, $E_u=E_i$ and $E_v=E_j$, and
applying Dirichlet boundary conditions, turns the right-hand side
into $Q_{ij}$.
More details about how such identities are found using a symbolic matrix
method will be postponed to a future publication~\cite{quasi}.

Finally, the other possibility is that $E_i=E_j$ (which need not imply
$i=j$).
We take Lemma~\ref{lem:overlap2d} with the choices
$\Omega_A=\Omega$, $u=\phi_i$, $v=\phi_j$, $E_i=E_j=E$, apply Dirichlet
boundary conditions, and use orthonormality (\ref{eq:on}). Thus
Lemma~\ref{lem:quasi} is proved for all choices of $i,j$.



\begin{thebibliography}{99}

\bibitem{agam}
O.~Agam and S.~Fishman, ``Semiclassical criterion for scars of wavefunctions
in chaotic systems'', Phys. Rev. Lett. {\bf 73} 806--809 (1994).

\bibitem{aurich}
R.~Aurich and M.~Taglieber, ``On the rate of quantum ergodicity
on hyperbolic surfaces and for billiards'',
Physica D {\bf 118}, 84--102 (1998).

\bibitem{austin}
E.~J.~Austin and M.~Wilkinson, ``Distribution of matrix elements of a
classicaly chaotic system'',
Europhys. Lett. {\bf 20}, 589--593 (1992).

\bibitem{baeckerbb}
A.~B\"{a}cker, R.~Schubert, and P.~Stifter,
``On the number of bouncing ball modes in billiards'',
J. Phys. A, {\bf 30}, 6783--95 (1997).

\bibitem{baecker}
A.~B\"{a}cker, R.~Schubert, and P.~Stifter,
``Rate of quantum ergodicity in Euclidean billiards'',
Phys. Rev. E {\bf 57}, 5425--47 (1998); also see Errata for this paper,
Phys. Rev. E {\bf 58} (4) (1998).

\bibitem{baeckerbim}
A.~B\"{a}cker,
``Numerical aspects of eigenvalue and eigenfunction computations
for chaotic quantum systems'',
in {\em The Mathematical Aspects of Quantum Maps}, M. Degli Esposti
and S. Graffi (Eds.) Springer Lecture Notes in Physics 618, 91-144 (2003).

\bibitem{mythesis}
A. H. Barnett,
Ph.~D.\ thesis, Harvard University, 2000.

\bibitem{dil}
A. H. Barnett, D. Cohen, and E. J. Heller,
Phys. Rev. Lett. {\bf 85}, 1412 (2000).

\bibitem{wlf}
A. H. Barnett, D. Cohen, and E. J. Heller,
~Rate of energy absorption for a driven chaotic cavity'',
J. Phys. A {\bf 34}, 413--437 (2001).

\bibitem{quasi}
A.~H.~Barnett,
``Quasi-orthogonality on the boundary for Euclidean Laplace eigenfunctions'',
preprint, {\tt math-ph/0601006}

\bibitem{scaling}
A.~H.~Barnett,
``The scaling method for the Dirichlet eigenproblem'', in preparation.

\bibitem{berry77}
M.~V.~Berry, ``Regular and irregular semiclassical wavefunctions'',
J. Phys. A {\bf 10} 2083--91 (1977).

\bibitem{berryscar}
M.~V.~Berry, in ``Les Houches Lecture Notes, Summer School on Chaos and
Quantum Physics'' (M.-J. Giannoni, A. Voros, and J. Zinn-Justin, Eds.),
Elsevier Science, Amsterdam, 1991;
M.~V.~Berry, Proc. Roy. Soc. A {\bf 243} 219 (1989).

\bibitem{mps}
T.~Betcke and L.~N.~Trefethen,
``Reviving the Method of Particular Solutions'',
SIAM Review {\bf 47}, 469--491 (2005).

\bibitem{bogscar}
E.~B.~Bogomolny, ``Smoothed wave functions of chaotic quantum systems'',
Physica D {\bf 31}, 169--189 (1988).

\bibitem{bogarith}
E.~B.~Bogomolny, B.~Georgeot, M.~Giannoni, and C.~Schmidt,
``Chaotic billiards generated by arithmetic groups'',
Phys. Rev. Lett. {\bf 69}, 1477--80 (1992).

\bibitem{RMT}
O.~Bohigas, ``Random matrix theories and chaotic dynamics'', in
{\em Chaos et physique quantique} (Les Houches, 1989), 87--199,
M. J. Giannoni, A. Voros, and J. Zinn-Justin, eds,
(North-Holland, Amsterdam, 1991).

\bibitem{boose}
D.~Boos\'{e} and J.~Main, ``Distributions of transition matrix elements in
classically mixed quantum systems'',
Phys. Rev. E {\bf 60}, 2813--44 (1999).

\bibitem{casati}
G.~Casati and T.~Prosen,
``The quantum mechanics of chaotic billiards'',
Physica D {\bf 131}, 293--310 (1999).

\bibitem{doronfrc}
D.~Cohen, ``Chaos and energy spreading for time-Dependent Hamiltonians,
and the various Regimes in the theory of Quantum Dissipation'',
Ann. Phys. (N.Y.) {\bf 283}, 175--231 (2000).


\bibitem{cdv}
Y.~Colin de Verdi\`{e}re,
``Ergodicit\'{e} et fonctions propres du laplacien'',
Comm. Math. Phys. {\bf 102} 497 (1985).


\bibitem{combe2}
M. Combescure and D. Robert,
``Semiclassical sum rules and generalized coherent states'',
J. Math. Phys. {\bf 36}, 6596--6610 (1995).

\bibitem{donnelly}
H.~Donnelly, ``Quantum Unique Ergodicity'', Proc. Am. Math. Soc.
{\bf 131}, 2945--51 (2003).

\bibitem{EFKAMM}
B.~Eckhardt, S.~Fishman, J.~Keating, O.~Agam, J.~Main, and K.~M\"{u}ller,
``Approach to ergodicity in quantum wave functions'',
Phys. Rev. E {\bf 52}, 5893--5903 (1995).

\bibitem{faure}
F.~Faure, S.~Nonnenmacher, and S.~De Bi\`{e}vre,
``Scarred eigenstates for quantum cat maps of minimal periods'',
Comm. Math. Phys. {\bf 239} 449--492 (2003).

\bibitem{fp86}
M.~Feingold and A.~Peres,
``Distribution of matrix elements of chaotic systems'',
Phys. Rev. A, {\bf 34}, 591 (1986).

\bibitem{stoc}
C.~V.~Gardiner,
{\em Handbook of Stochastic Methods for physics, chemistry,
and the natural sciences}, 2nd Ed.\ 
(Springer-Verlag, 1997).

\bibitem{gutz}
M.~C.~Gutzwiller,
{\em Chaos in Classical and Quantum Mechanics},
(Springer, NY, 1990).

\bibitem{hel84}
E.~J.~Heller, ``Bound-state eigenfunctions of classically chaotic
hamiltonian systems: scars of periodic orbits'',
Phys. Rev. Lett. 1515--18 (1984).

\bibitem{hellerhouches}
E. J. Heller, ``Wavepacket dynamics and quantum chaology'',
in {\em Chaos et physique quantique} (Les Houches, 1989), 547--664,
M. J. Giannoni, A. Voros, and J. Zinn-Justin, eds,
(North-Holland, Amsterdam, 1991).


\bibitem{hort}
S.~Hortikar and M.~Srednicki,
``Random matrix elements and eigenfunctions in chaotic systems'',
Phys. Rev. E {\bf 57}, 7313--16 (1998).

\bibitem{KS}
J.~R.~Kuttler and V.~G.~Sigillito, ``Eigenvalues of the Laplacian in
two dimensions'', SIAM Review, {\bf 26}, 163--193 (1984).

\bibitem{li}
B.~Li and B.~Hu, J. Phys. A {\bf 31}, 483 (1998).

\bibitem{linden}
E.~Lindenstrauss,
``Invariant measures and arithmetic quantum unique ergodicity'',
Annals of Math. (to appear).


\bibitem{lsquad}
W.~Luo and P.~Sarnak, ``Quantum variance for Hecke eigenforms'',
Annales Scient. de l'\'{E}cole Norm. Sup. {\bf 37}, 769--799 (2004).

\bibitem{martinez}
A.~Martinez,
{\em An Introduction to Semiclassical and Microlocal Analysis},
(Springer, 2002).

\bibitem{lev}
L.~Kaplan and E.~J.~Heller, ``Linear and nonlinear theory of eigenfunction
scars'',
Ann. Phys. {\bf 264}, 171--206 (1998).

\bibitem{scarmometer}
L.~Kaplan and E.~J.~Heller, ``Measuring scars of periodic orbits'',
Phys. Rev. E {\bf 59}, 6609--28 (1999).


\bibitem{prosen}
T.~Prosen,
``Statistical properties of matrix elements in a Hamilton system between
integrability and chaos'',
Ann. Phys. N.Y. {\bf 235}, 115--164 (1994).

\bibitem{prosen3d}
T.~Prosen,
``Quantization of a generic chaotic 3D billiard with smooth boundary.
I. Energy level statistics'', Phys. Lett. A {\bf 233}, 323--331 (1997);
T.~Prosen,
``Quantization of a generic chaotic 3D billiard with smooth boundary.
II. Structure of high-lying eigenstates'', Phys. Lett. A {\bf 233},
332--342 (1997).

\bibitem{RS}
Z.~Rudnick and P.~Sarnak, ``The behaviour of eigenstates of arithmetic
hyperbolic manifolds'', Comm. Math. Phys. {\bf 161} 195--213 (1994).

\bibitem{sarnak95}
P.~Sarnak, ``Arithmetic quantum chaos'', The Schur lectures (Tel Aviv, 1992),
Israel Math. Conf. Proc. {\bf 8}, 183--236 (Bar-Ilan University, 1995).

\bibitem{sarnakCRM}
P.~Sarnak, ``Spectra and eigenfunctions of Laplacians'', in
{\em Partial Differential Equations and Their Applications (CRM
Proceedings and Lecture Notes, Vol 12)}
pp. 261--276 (American Mathematical Society, 1997).

\bibitem{baltimore}
P.~Sarnak, ``Spectra of Hyperbolic Surfaces'', Bulletin of the AMS,
{\bf 40}, 441--478 (2003).

\bibitem{schanz}
H.~Schanz and T.~Kottos, ``Scars on quantum networks ignore the
Lyapunov exponent'', Phys. Rev. Lett. {\bf 90}, 234101 (2003).

\bibitem{schnir}
A.~I.~Schnirelman, Usp. Mat. Nauk. {\bf 29}, 181 (1974).

\bibitem{disperse}
Ya.~G.~Sinai, ``Dynamical systems with elastic reflections:
ergodic properties of dispersing billiards'',
Russ. Math. Surveys {\bf 25}, 137--189 (1970).

\bibitem{bayes}
D.~S.~Sivia, {\em Data Analysis: A Bayesian Tutorial}, (Oxford, 1996).

\bibitem{history}
F.~Steiner, ``Quantum Chaos'', {\tt chao-dyn/9402001} (1994).

\bibitem{tanner}
G.~Tanner,
``How chaotic is the stadium billiard? A semiclassical analysis'',
J. Phys. A {\bf 30}, 2863--88 (1997).

\bibitem{tate}
T. Tate, 
``Some remarks on the off-diagonal asymptotics in quantum ergodicity'',
Asym. Anal. {\bf 19}, 289--296 (1999).

\bibitem{scalinguse1}
G.~Veble, M.~Robnik, and J.~Liu,
``Study of regular and irregular states in generic systems'',
J. Phys. A {\bf 32}, 6423--44 (1999).

\bibitem{v+s}
E. Vergini and M. Saraceno,
Phys. Rev. E, {\bf 52}, 2204 (1995); E. Vergini,
Ph.~D.\ thesis, Universidad de Buenos Aires, 1995.

\bibitem{wilk87}
M.~Wilkinson, ``A semiclassical sum rule for matrix elements of
classically chaotic systems'', J. Phys. A {\bf 20}, 2415--23 (1987).

\bibitem{zel}
S.~Zelditch, ``Uniform distribution of eigenfunctions on compact
hyperbolic surfaces'', Duke Math. J. {\bf 55}, 919 (1987).

\bibitem{zeloffdiag}
S.~Zelditch, ``Quantum transition amplitudes for ergodic and for
completely integrable systems'',
J. Funct. Anal. {\bf 94}, 415--436 (1990).

\bibitem{zonb1}
S.~Zelditch, ``Quantum ergodicity on the sphere'',
Comm. Math. Phys. {\bf 146}, 61--71 (1992).

\bibitem{zelupper}
S.~Zelditch, ``On the rate of quantum ergodicity I: Upper Bounds'',
Comm. Math. Phys. {\bf 160}, 81--92 (1994).

\bibitem{zellower}
S.~Zelditch, ``On the rate of quantum ergodicity II: Lower Bounds'',
Comm. Partial Diff. Eqns. {\bf 19}, 1565--79 (1994).

\bibitem{zqmix}
S.~Zelditch, ``Quantum mixing'', J. Func. Anal. {\bf 140}, 68--86 (1996).

\bibitem{zonb2}
S.~Zelditch, ``A random matrix model for quantum mixing'',
Int. Math. Res. Not. {\bf 3} 115--137 (1996).

\bibitem{zzw}
S.~Zelditch and M.~Zworski, ``Ergodicity of eigenfunctions for
ergodic billiards'', Comm. Math. Phys., {\bf 175} 673--682 (1996).

\bibitem{zencyc}
S.~Zelditch, ``Quantum ergodicity and mixing of eigenfunctions'',
Elsevier Encyclopedia of Mathematical Physics, preprint (2005),
{\tt math-ph/0503026}





\end{thebibliography}
\end{document}